\documentclass[11pt,a4paper]{article}

\RequirePackage{ifpdf} 
\usepackage{amsmath} 
\usepackage{mathtools}

\usepackage{jheppub}
\usepackage{pstricks}
\usepackage[final]{pdfpages} 
\usepackage{ifpdf} 
\usepackage{slashed}

\usepackage{hyperref}

\usepackage{color} 
\usepackage{graphics}

\usepackage{etoolbox} 
\usepackage{fixmath}

\usepackage{caption} 
\usepackage{subcaption} 
\usepackage{amsfonts}

\usepackage{multirow}
\usepackage{epstopdf}

\usepackage{relsize}
\usepackage{float}

\usepackage{tikz}
\usetikzlibrary{positioning,arrows}
\usetikzlibrary{decorations.pathmorphing}
\usetikzlibrary{decorations.markings}
\usetikzlibrary{shapes.geometric}
\tikzset{
    vector/.style={decorate, decoration={snake}, draw},
    provector/.style={decorate, decoration={snake,amplitude=2.5pt}, draw},
    antivector/.style={decorate, decoration={snake,amplitude=-2.5pt}, draw},
    fermion/.style={draw=black,
      postaction={decorate},decoration={markings,mark=at position .55
        with {\arrow[draw=black]{>}}}}, 
    fermionbar/.style={draw=black, postaction={decorate},
                       decoration={markings,mark=at position .55 with {\arrow[draw=black]{<}}}},
    fermionnoarrow/.style={draw=black},
    gluon/.style={decorate, draw=black,decoration={coil,amplitude=4pt, segment length=4pt}},
    scalar/.style={dashed,draw=black,
      postaction={decorate},decoration={markings,mark=at position .55
        with {\arrow[draw=black]{>}}}}, 
    scalarbar/.style={dashed,draw=black,
      postaction={decorate},decoration={markings,mark=at position .55
        with {\arrow[draw=black]{<}}}}, 
    scalarnoarrow/.style={dashed,draw=black},
    electron/.style={draw=black,
      postaction={decorate},decoration={markings,mark=at position .55
        with {\arrow[draw=black]{>}}}}, 
    bigvector/.style={decorate, decoration={snake,amplitude=4pt}, draw},
}

\title{NNLO QCD corrections to production of a spin-2 particle with
non-universal couplings in the DY process}

\author[a,b]{Pulak Banerjee,}
\author[a,b]{Prasanna K. Dhani,}
\author[c]{M.C. Kumar,}
\author[d,b]{Prakash Mathews,} 
\author[a,b]{and V. Ravindran}

\affiliation[a]{The Institute of Mathematical Sciences, IV Cross Road,
  CIT Campus, Chennai 600 113, Tamil Nadu, India}
\affiliation[b]{Homi Bhabha National Institute, Training School Complex, Anushakti Nagar, Mumbai 400085, India}
\affiliation[c]{Department of Physics, Indian Institute of Technology Guwahati, Guwahati 781039,
  India} 
\affiliation[d]{Saha Institute of Nuclear Physics, 1/AF Bidhan Nagar,
  Kolkata 700 064, West Bengal, India}

\emailAdd{bpulak@imsc.res.in}
\emailAdd{prasannakd@imsc.res.in} \emailAdd{mckumar@iitg.ac.in}
\emailAdd{prakash.mathews@saha.ac.in} 
\emailAdd{ravindra@imsc.res.in}

\abstract{
We study the phenomenological impact of the interaction of spin-2 fields 
with those of the Standard Model in a model independent framework up to next-to-next-to-leading order
in perturbative Quantum Chromodynamics. 
We use the invariant mass distribution of the pair of leptons produced at the Large Hadron Collider to demonstrate this.
A minimal scenario where the spin-2 fields couple 
to two gauge invariant operators with different coupling strengths has been considered.  
These operators being not conserved show very different ultraviolet behaviour increasing 
the searches options of spin-2 particles at the colliders.  
We find that our results using the higher order quantum corrections stabilise the
predictions with respect to renormalisation and factorisation scales. 
We also find that corrections are appreciable which need to be taken into account in such searches
at the colliders.
}

\preprint{} 

\begin{document}
\allowdisplaybreaks[4]
\unitlength1cm
\maketitle
\flushbottom

%************
% Definition
%************

\def\D{{\cal D}}
\def\DD{\overline{\cal D}}
\def\g{\overline{\cal G}}
\def\gm{\gamma}
\def\M{{\cal M}}
\def\ep{\epsilon}
\def\epm1{\frac{1}{\epsilon}}
\def\epm2{\frac{1}{\epsilon^{2}}}
\def\epm3{\frac{1}{\epsilon^{3}}}
\def\epm4{\frac{1}{\epsilon^{4}}}
\def\unM{\hat{\cal M}}
\def\ashat{\hat{a}_{s}}
\def\asmur{a_{s}^{2}(\mu_{R}^{2})}
\def\sigbar{{{\overline {\sigma}}}\left(a_{s}(\mu_{R}^{2}), L\left(\mu_{R}^{2}, m_{H}^{2}\right)\right)}
\def\sigbarn{{{{\overline \sigma}}_{n}\left(a_{s}(\mu_{R}^{2}) L\left(\mu_{R}^{2}, m_{H}^{2}\right)\right)}}
\def\sigh{\hat{\sigma}}
\def\unas{ \left( \frac{\hat{a}_s}{\mu_0^{\epsilon}} S_{\epsilon} \right) }
\def\rnM{{\cal M}}
\def\bt{\beta}
\def\cD{{\cal D}}
\def\cC{{\cal C}}
\def\ca{\text{\tiny C}_\text{\tiny A}}
\def\cf{\text{\tiny C}_\text{\tiny F}}
\def\ct{{\red []}}
\def\sv{\text{SV}}
\def\murOmu{\left( \frac{\mu_{R}^{2}}{\mu^{2}} \right)}
\def\bb{b{\bar{b}}}
\def\bt0{\beta_{0}}
\def\bt1{\beta_{1}}
\def\bt2{\beta_{2}}
\def\bt3{\beta_{3}}
\def\gm0{\gamma_{0}}
\def\gm1{\gamma_{1}}
\def\gm2{\gamma_{2}}
\def\gm3{\gamma_{3}}
\def\eps{\epsilon}
\def\l{\left}
\def\r{\right}

%\newcommand{\dis}[1]{\color{blue}\mathbold{#1}}
%************
% Definition
%************
\def\del{{\partial}}
\def\D{{\cal D}}
\def\DD{\overline{\cal D}}
\def\g{\overline{\cal G}}
\def\gm{\gamma}
\def\M{{\cal M}}
\def\ep{\epsilon}
\def\epm1{\frac{1}{\epsilon}}
\def\epm2{\frac{1}{\epsilon^{2}}}
\def\epm3{\frac{1}{\epsilon^{3}}}
\def\epm4{\frac{1}{\epsilon^{4}}}
\def\unM{\hat{\cal M}}
\def\ashat{\hat{a}_{s}}
\def\asmur{a_{s}^{2}(\mu_{R}^{2})}
\def\sigbar{{{\overline {\sigma}}}\left(a_{s}(\mu_{R}^{2}), L\left(\mu_{R}^{2}, m_{H}^{2}\right)\right)}
\def\sigbarn{{{{\overline \sigma}}_{n}\left(a_{s}(\mu_{R}^{2}) L\left(\mu_{R}^{2}, m_{H}^{2}\right)\right)}}
\def\sigh{\hat{\sigma}}
\def\unas{ \left( \frac{\hat{a}_s}{\mu_0^{\epsilon}} S_{\epsilon} \right) }
\def\rnM{{\cal M}}
\def\bt{\beta}
\def\cD{{\cal D}}
\def\cC{{\cal C}}
\def\ca{\text{\tiny C}_\text{\tiny A}}
\def\cf{\text{\tiny C}_\text{\tiny F}}
\def\ct{{\red []}}
\def\sv{\text{SV}}
\def\murOmu{\left( \frac{\mu_{R}^{2}}{\mu^{2}} \right)}
\def\bb{b{\bar{b}}}
\def\bt0{\beta_{0}}
\def\bt1{\beta_{1}}
\def\bt2{\beta_{2}}
\def\bt3{\beta_{3}}
\def\gm0{\gamma_{0}}
\def\gm1{\gamma_{1}}
\def\gm2{\gamma_{2}}
\def\gm3{\gamma_{3}}
\def\l{\left}
\def\r{\right}

\newcommand{\dis}{}
\newcommand{\overbar}[1]{mkern-1.5mu\overline{\mkern-1.5mu#1\mkern-1.5mu}\mkern
1.5mu}

\newcommand{\nn}{\nonumber\\}
\newcommand{\be}{\begin{equation}}
\newcommand{\ee}{\end{equation}}
\newcommand{\bea}{\begin{eqnarray}}
\newcommand{\eea}{\end{eqnarray}}

%%%%%
%%%%%

%*******
% Intro
%*******
\section{Introduction}
%%%% Start%%%%

With the absence of any signal of new physics at the large hadron
collider (LHC) at present energies, searches of physics beyond the
Standard Model (BSM) is based on the ability to make very precise
theoretical predictions within the Standard Model (SM) and to look
for possible deviations between experimental observations and
theoretical predictions, as a hint of new physics, within estimated
uncertainties.  In order to constrain the new physics model
parameters, one needs to also compute the BSM signals to the same level
of theoretical precision as the SM and compare with the observations
made at the LHC.  Quantum Chromodymanics (QCD) corrections are large at the LHC and inclusion of
higher order terms reduces the theoretical uncertainties substantially.
Many SM processes have been measured at the LHC and have cross sections
that are in excellent agreement with higher order QCD predictions.  
This has helped
in the discovery of the Higgs boson by ATLAS \cite{Aad:2012tfa} and CMS
\cite{Chatrchyan:2012xdj}
collaborations at the LHC and hence the measurement of the important fundamental
parameter of the SM, the Higgs mass $m_H$ (see 
\cite{Harlander:2002wh,Anastasiou:2002yz,Ravindran:2003um}). 
Precise measurement of the
Higgs mass is essential for the understanding of the stability of
electroweak vacuum \cite{Degrassi:2012ry}.

In spite of the fact that the SM is in excellent agreement with 
experimental observations, we know that there are compelling
reasons to go beyond the SM.  In the context of the discovery
of a boson at 125 GeV in the di-photon channel, models with
spin-2 were also necessary to ascertain the spin and parity of
the discovered boson.  In the mean time the bounds on conventional models
such as the Randall-Sundrum models with warped extra dimensions~\cite{Randall:1999ee},
 where
the spin-2 couples universally
to the SM energy momentum tensor was much higher. 
A universally-coupled spin-2 particle is heavily constrained
\cite{Khachatryan:2016yec,Aaboud:2017eta}.
Models
with non-universal coupling of a spin-2 to SM was hence
a suitable alternative.  In this model, the spin-2 couples to, two
sets of gauge invariant SM tensorial operators with different
coupling strengths, but are
not individually conserved.  The universal coupling would correspond
to the coupling strength being equal and the tensorial operators
adding up to the conserved energy momentum tensor.  Models
with non-universal coupling were incorporated in tools like
Higgs charaterisation \cite{Artoisenet:2013puc} to NLO in QCD.  
Non universal coupling 
lead to additional challenges: (a) additional UV renornalisation were
needed, (b) in the IR sector, additional double and single pole
terms had to be cancelled with the counter parts from real emission 
process and mass factorisation counter terms,  thus demonstrating
the IR factorisation to NLO for non-universal coupling \cite{Artoisenet:2013puc}.
Note that we take this for granted in perturbative QCD (pQCD) and for universal coupling it is guaranteed
by the conserved energy-momentum tensor. 

Recently, the UV structure of non-universal coupling up to three
loop order in QCD was investigated \cite{Ahmed:2016qjf} where in the spin-2
fields couple to two sets of gauge invariant tensorial operators
constructed out of the SM fields (with different coupling strengths).
These rank-2 operators are unfortunately not conserved, unlike
energy-momentum tensor of QCD~\cite{Nielsen:1977sy}. 
Consequently, both these operators as well as the couplings get
additional UV renormalisation order by order in perturbation theory.
Exploiting the universal IR structure of QCD amplitudes
even in the case of a non-universal spin-2 coupling, on-shell
form factors of these operators between quark and gluon states
have been computed.  These are important ingredients for
observables at the LHC, to study models with such interactions.

For universal coupling, depending on the geometry of extra dimensions,
{\em viz}.\ large extra dimensions or warped extra dimension models,
studies have been extensively carried out upto higher orders in QCD in various
channels that are relevant for the LHC.  In these models, the DY process
has been studied to NLO \cite{Mathews:2004xp,Mathews:2005zs,Kumar:2006id}
for various observables.  Di-vector boson final
state have been studied to NLO level in
\cite{Kumar:2008pk, Kumar:2009nn, Agarwal:2009xr, Agarwal:2009zg,
Agarwal:2010sp, Agarwal:2010sn}.
To NLO+parton shower (PS) accuracy all the non-color, di-final states have been studied
\cite{Frederix:2012dp, Frederix:2013lga, Das:2014tva} in the aMC@NLO
framework.  
Production of a generic spin-2 particle in association with coloured
particles, vector bosons and the Higgs boson have been studied in
\cite{Das:2016pbk} to NLO+PS accuracy.  To the next higher order in QCD
the form factor of a spin-2 universally coupled to quarks and gluons up
to two loops was computed in \cite{deFlorian:2013sza}.  Subsequently the
next-to-next-to-leading order (NNLO) computation in the threshold limit was done in \cite{deFlorian:2013wpa}
and finally the full NNLO computation in \cite{Ahmed:2016qhu}.
Production of a spin-2 in association with a jet to full two-loop QCD
corrections has also been completed recently with the evaluation of
generic spin-2 decaying to $g~ g~ g$ \cite{Ahmed:2014gla} and 
$ q ~\bar q ~g$ \cite{Ahmed:2016yox}.

The di-lepton final state is the most studied and a very clean
process at the LHC. In BSM scenarios the dilepton signal could 
be enhanced due to additional contributions from BSM intermediate
states that could couple to a di-lepton.  For the universal
spin-2 coupling the DY
process has been evaluated upto NNLO in QCD.  This involved
various steps: to begin with NLO corrections were evaluated
\cite{Mathews:2004xp}, followed by the two loop quark and
gluon form factors \cite{deFlorian:2013sza}, which lead to the
computation of NNLO QCD corrections to the graviton
production in models of TeV-scale gravity, within the soft-virtual
approximation \cite{deFlorian:2013wpa}.  Finally the complete NNLO QCD corrections to the
production of di-leptons at hadron colliders in large extra
dimension models with spin-2 particles are reported in
\cite{Ahmed:2016qhu}.  

%With the recent progress in precision study 
%of non-universal spin-2 coupling, in this article we look
%at its phenomenological implications at the LHC.
The non-universal coupling of spin-2 to SM has been actively
considered by the ATLAS Collaboration \cite{Aad:2015mxa,Pedersen:2015jdh} to provide
exclusion of several non-SM spin hypotheses.  This analysis
has been done in the Higgs characterisation frame work \cite{Artoisenet:2013puc,Das:2016pbk} 
to NLO+PS accuracy.  With the recent results \cite{Ahmed:2016qjf} upto three loop form factors of a massive spin-2
particle with non-universal coupling, NNLO computation is now
possible.  In this article we look at the phenomenological
implications of these models to NNLO at the LHC.

The paper is organised as follows. We discuss the effective action that
describes how spin-2 particle couples to those of the SM through two gauge
invariant operators with renormalisable coupling. Using this action, we
compute QCD radiative correction to the production of pair of leptons in
particular their invariant mass distribution up to NNLO level. A detailed
phenomenological study on the impact of our results is presented for the
LHC. Finally we conclude. The relevant form factors are presented in the appendix 
and mass factorised partonic cross sections are given as electronically readable version.

\section{Theoretical Framework}
\label{sec:Theory}

\subsection{Effective action}

The interaction part of the effective action describes the non-universal coupling of
the spin-2 fields denoted by $h_{\mu\nu}$ with those of QCD, consists of
two gauge invariant operators, namely $\hat {\cal O}^G_{\mu\nu}$ and 
$\hat {\cal O}^Q_{\mu\nu}$ and is given by 
\begin{equation}
\label{eq:action}
S = -\frac{1}{2} \int d^4x ~ h^{\mu\nu}(x) \left(\hat \kappa_G~ \hat {\cal O}^{G}_{\mu\nu} (x) 
	+ \hat \kappa_Q~\hat {\cal O}^{Q}_{\mu\nu}(x) \right) \,,
\end{equation}
where $\hat \kappa_{G,Q}$ are dimension full couplings, the pure gauge
sector is denoted by $G$, while $Q$ denotes the fermionic sector and
its gauge interaction.  This decomposition is not unique as one can
adjust gauge invariant terms between them.  The gauge invariant
operators $\hat {\cal O}^G_{\mu\nu}$ and $\hat {\cal O}^Q_{\mu\nu}$
are as follows:
\begin{eqnarray} \label{eq:emtensor}
\hat {\cal O}^G_{\mu\nu} &=&{1 \over 4} g_{\mu\nu} \hat F_{\alpha \beta}^a \hat F^{a\alpha\beta} 
- \hat F_{\mu\rho}^a \hat F^{a\rho}_\nu
 - \frac{1}{\hat \xi} g_{\mu\nu} \partial^\rho(\hat A_\rho^a\partial^\sigma \hat A_\sigma^a)
-{1 \over 2\hat \xi}g_{\mu\nu} \partial_\alpha \hat A^{a\alpha} \partial_\beta \hat A^{a\beta} 
 \nonumber\\
&& + \frac{1}{\hat \xi}(\hat A_\nu^a \partial_\mu(\partial^\sigma \hat A_\sigma^a) + \hat A_\mu^a\partial_\nu
 (\partial^\sigma \hat A_\sigma^a))
+\partial_\mu \overline {\hat \omega^a} (\partial_\nu \hat \omega^a - \hat g_s f^{abc} \hat A_\nu^c \hat \omega^b)
\nonumber\\
&& +\partial_\nu \overline {\hat \omega^a} (\partial_\mu \hat \omega^a- \hat g_s f^{abc} \hat A_\mu^c \hat \omega^b)
-g_{\mu\nu} \partial_\alpha \overline {\hat \omega^a} (\partial^\alpha \hat \omega^a - \hat g_s f^{abc} \hat A^{c \alpha} \hat \omega^b) \,,
\\
 \hat {\cal O}^Q_{\mu\nu} &= &
 \frac{i}{4} \Big[ \overline {\hat \psi} \gamma_\mu (\overrightarrow{\partial}_\nu -i \hat g_s T^a \hat A^a_\nu)\hat \psi
 -\overline {\hat \psi} (\overleftarrow{\partial}_\nu + i \hat g_s T^a \hat A^a_\nu) \gamma_\mu \hat \psi
 +\overline {\hat \psi} \gamma_\nu (\overrightarrow{\partial}_\mu -i \hat g_s T^a \hat A^a_\mu)\hat \psi
 \nonumber\\
 &&-\overline {\hat \psi} (\overleftarrow{\partial}_\mu + i \hat g_s T^a \hat A^a_\mu) \gamma_\nu \hat \psi\Big]
- ig_{\mu\nu} \overline {\hat \psi} \gamma^\alpha (\overrightarrow{\partial}_\alpha -i \hat g_s T^a \hat A^a_\alpha)\hat \psi \,,
 \end{eqnarray}
in the above equations the unrenormalised quantities are denoted by hat $( ~{\hat{}} ~)$.
$\hat g_s$ is the strong coupling constant, ${\hat \xi}$ the gauge fixing parameter,
$\hat A_\nu^c$ the gauge field, ${\hat \psi}$ the quark field and ${\hat \omega^a}$ the
ghost fields.  The structure constants of $SU(N)$ gauge group are denoted by $f^{abc}$ and
the Gell-Mann matrices by $T^a$.
The sum of $\hat{{ \cal O}}_G$ and $\hat{{\cal O}}_Q$ is the energy momentum tensor of the QCD
part and is protected by radiative corrections to all orders, thanks to fact that
it is conserved.  The Feynman rules for the non-universal case in contrast to the
universal case \cite{Han:1998sg,Mathews:2004pi}, would have a
prefactor $\kappa_Q$ for the coupling for a spin-2 to a pair of fermions or any fermionic 
SM vertex, while a spin-2 coupling to gluons, ghosts or any SM gauge or ghost vertex 
would have a prefactor $\kappa_G$.  The individual gauge ${\cal O}_G$ and fermionic
${\cal O}_Q$ operators are not conserved in QCD and hence require additional ultraviolet
(UV) counter terms
in order to renormalise them.  In \cite{Ahmed:2016qjf}, we determined these additional UV
renormalisation constants up to three loop level in QCD.  We obtained them by exploiting
the universal infrared properties of on-shell amplitudes involving these composite
operators.  Since we have two operators at our disposal, they mix under renormalisation
as follows:
\begin{equation}
  \label{eq:Zmat}
%\[
\begin{bmatrix}
O^{G} \\ O^{Q}
\end{bmatrix}
=
\begin{bmatrix}
  Z_{GG} & Z_{GQ} \\ Z_{QG} & Z_{QQ}
\end{bmatrix}
\begin{bmatrix}
\hat  O^{G} \\ \hat O^{Q}
\end{bmatrix}
\,.
%\]
\end{equation}
where the renormalisation constants $Z_{IJ}$ in terms of the anomalous dimensions
$\gamma_{IJ} = \sum_{n=1}^{\infty} a_{s}^{n} \gamma_{IJ}^{(n)}$ are given by
\begin{align}
  Z_{IJ} &= \delta_{IJ} 
           + {a}_{s} \Bigg[ \frac{2}{\epsilon}
           \gamma_{IJ}^{(1)} \Bigg] 
           + {a}_{s}^{2} \Bigg[
           \frac{1}{\epsilon^{2}} \Bigg\{  2
           \beta_{0} \gamma_{IJ}^{(1)} + 2 \gamma_{IK}^{(1)} \gamma_{KJ}^{(1)}  \Bigg\} + \frac{1}{\epsilon} \Bigg\{ \gamma_{IJ}^{(2)}\Bigg\}
           \Bigg]  \,,
\end{align}
where $I,J=G,Q$, $a_s \equiv g_s^2/16 \pi^2$ and space-time dimension is taken to be $d=4+\epsilon$.
The renormalisation constants $Z_{IJ}$ computed in 
\cite{Ahmed:2016qjf} are given below up to $a_s^2$ for completeness:
\begin{align}
Z_{GG} &=  1 + a_s\left[ -\frac{4}{3\epsilon}n_f \right]+a_s^2\bigg[ \frac{1}{\epsilon^2}\left\{ -\frac{44}{9}C_A n_f + \frac{32}{9}C_F n_f +\frac{16}{9}n_f^2\right\}
\nonumber\\&
+\frac{1}{\epsilon}\left\{-\frac{35}{27}C_A n_f - \frac{74}{27}C_F n_f\right\} \bigg]\,,
\nonumber\\
Z_{GQ}&= a_s \left[\frac{16}{3\epsilon}C_F \right] + a_s^2\bigg[\frac{1}{\epsilon^2}\left\{\frac{176}{9}C_A C_F -\frac{64}{9}C_F n_f -\frac{128}{9}C_F^2 \right\} 
\nonumber\\&
+ \frac{1}{\epsilon}\left\{\frac{376}{27}C_A C_F -\frac{104}{27}C_F n_f -\frac{112}{27}C_F^2 \right\} \bigg]\,,
\nonumber\\
Z_{QG}&= a_s\left[\frac{4}{3\epsilon} n_f\right] + a_s^2\bigg[ \frac{1}{\epsilon^2}\left\{\frac{44}{9}C_A n_f -\frac{32}{9}C_F n_f -\frac{16}{9} n_f^2 \right\} 
\nonumber\\&
+\frac{1}{\epsilon}\left\{ \frac{35}{27}C_A n_f +\frac{74}{27} C_F n_f\right\}\bigg]\,,
\nonumber\\
Z_{QQ}&= 1+ a_s\left[-\frac{16}{3\epsilon} \right] + a_s^2\bigg[\frac{1}{\epsilon^2}\left\{-\frac{176}{9}C_A C_F +\frac{64}{9}C_F n_f +\frac{128}{9}C_F^2 \right\}
\nonumber\\&
+ \frac{1}{\epsilon}\left\{-\frac{376}{27}C_A C_F + \frac{104}{27}C_F n_f +\frac{112}{27}C_F^2 \right\}\bigg]\,,
\end{align}
where $C_A=N$ and $C_F=(N^2-1)/2N$ are the quadratic Casimirs of the $SU(N)$ group and $n_f$
is the number of quark flavours.  The fact that the energy momentum tensor 
$T_{\mu \nu} = {\cal O}^G_{\mu\nu} + {\cal O}^Q_{\mu\nu}$ is conserved leads to 
$\gamma^{(n)}_{QG}=-\gamma^{(n)}_{GG}$ and $\gamma^{(n)}_{QQ}=-\gamma^{(n)}_{GQ}$ or
equivalently $Z_{GG} = 1-Z_{QG}$ and $Z_{QQ}=1-Z_{GQ}$, which is expected to be true to all
orders in $a_s$.  
All $\gamma^{(n)}_{GG}$ are proportional to $n_f$ which is consistent with the
expectation that the conserved property of ${\cal O}^G_{\mu\nu}$ breaks down beyond
tree level due to the presence of quark loops.   For pure gauge theory 
$(n_f=0)$ and the energy momentum tensor of the pure gauge theory 
${\cal O}^G_{\mu\nu}$ is hence conserved by itself.  

Defining the renormalised  $\kappa_I$ in terms of bare ones through 
$\hat \kappa_I = \sum_{J=G,Q} Z_{IJ} \kappa_J$ with
$I,J = G,Q$,
we find that the action takes the following form
\begin{eqnarray}
	S &=& -{1 \over 2} \int d^4x ~ h_{\mu\nu} \left(\kappa_G~ {\cal O}^{G,\mu\nu} + \kappa_Q~{\cal O}^{Q,\mu\nu} \right)\,,
\end{eqnarray}
the resulting interaction terms expressed in terms of renormalised operators and
renormalised couplings are guaranteed to predict UV finite quantities to all
orders in strong coupling.  In the rest of the paper, we will use this version of
the Lagrangian to study the phenomenology.

% % % % % % % % % % % % % % % % % % % % % % % 
% % % % % % % % % % % % % % % % % % % % % % % 

% % % % % % % % % % % % % % % % % % % % % % % 

\subsection{Lepton pair invariant mass distribution $d\sigma/dQ^2$}
\label{ss:inv}
Our next task to use the effective action expressed in terms of renormalised operators ${\cal O}_I$ and couplings $\kappa_I$ to
obtain production cross section for a 
pair of leptons $(l^+,l^-)$, through the
scattering of two protons $H_1,H_2$ at the LHC:  
\begin{equation}
\label{eq:3}
H_1 (P_1)+ H_2 (P_2)  \rightarrow l^+(l_1) + l^-(l_2) + X(P_X)\,
\end{equation}
where the 4-momenta of the corresponding particles are denoted in the 
parentheses and the final inclusive state is denoted by $X$.
The hadronic cross section is related to the partonic subprocess 
cross sections in the QCD improved parton model as
\begin{align}
\label{eq:4}
2S \frac{d\sigma^{H_1H_2}}{d Q^2} \big(\tau, Q^2 \big)  = \sum_{ab =
  q,\bar{q},g} \int_{0}^{1} dx_1 \int_{0}^{1} dx_2 ~~  
	\hat f_a^{H_1}(x_1)
	\hat f_b^{H_2} (x_2)  
\nonumber\\
\times \int_0^1dz \, 2 s\frac{d\hat{\sigma}^{ab}}{dQ^2}\big(z,
  Q^2\big)\delta(\tau - zx_1x_2)\,, 
\end{align}
where $Q^2$ is the invariant mass square of the final state leptonic pair 
and $S$ is the square of the hadronic center of mass energy which is
related to the partonic one, $s$, through $s=x_{1}x_{2}S$, 
similarly $\tau \equiv Q^2/S$, $z \equiv Q^2/s$
and $\tau = x_1 x_2z$.  The unrenormalised partonic distribution functions of the
partons $a$ and $b$ are $\hat f_a$ and $\hat f_b$ respectively.
The partonic sub process corresponding to the hadronic process is
\begin{align*}
a(p_{1})+b(p_{2})
\rightarrow j(q) + \sum\limits_{i=1}^{m} X_{i}(q_{i})  \,,
%\rightarrow l^{+}(l_{1}) + l^{-}(l_{2}) + \sum\limits_{i=1}^{m} X_{i}(q_{i})
\end{align*}
where the summation over $i$ corresponds to all the real QCD final state
partons that could contribute to a particular order in perturbative QCD.  The 
initial state partons $a b \to j$, a  neutral state $j$ which could be a
photon ($\gamma^{*}$), Z-boson ($Z^*$) or spin-2 particle and further decays to pair of leptons
$j \to l^+ l^-$.   

At the partonic level, one encounters amplitudes involving both SM vector
bosons and spin-2 particles as propagators and hence, at the cross section
level, the squared amplitudes contain in addition to contributions from SM
and spin-2 separately, those from interference of SM and spin-2 amplitudes.
Interestingly, for the invariant mass distributions, the later one identically
vanishes for the universal case, which was earlier noted both at NLO and NNLO
levels in ~\cite{Mathews:2004xp,Ahmed:2016qhu}.  Hence, at the cross section
level, the SM and spin-2 contributions simply add up as
\begin{align}
2 S{d \sigma^{H_1H_2} \over dQ^2}(\tau,Q^2)&=
2 S{d \sigma^{H_1H_2}_{\rm SM} \over dQ^2}(\tau,Q^2)
+2 S{d \sigma^{H_1H_2}_{\rm spin-2} \over dQ^2}(\tau,Q^2) \,,
\end{align}
where the SM results are known exactly upto NNLO level for long time (see \cite{Altarelli:1978id, Matsuura:1987wt, Matsuura:1988sm,Hamberg:1990np}) and result at N$^3$LO in the 
soft gluon approximation is also available, see \cite{Ahmed:2014cla}.
For the spin-2 case with universal coupling, namely $\kappa_G=\kappa_Q=\kappa$, 
the results upto NNLO level can be found in \cite{Mathews:2004xp,Ahmed:2016qhu}. 
In this article, we have extended this
computation to NNLO QCD 
%from the existing NLO 
%result~\cite{Mathews:2004xp} in models with spin-2 particles,
%namely, the contributions coming from
for the case of non-universal couplings
i.e., when $\kappa_G$ and $\kappa_Q$ are different.
We briefly describe the methodology that we use to obtain
the mass factorised partonic cross sections up to NNLO level.
Unlike the SM, for the spin-2  exchange, at
leading order (LO) we can have
gluon initiated sub process in addition to the quark initiated
one: 
\begin{align}
\label{eq:16}
q+{\bar q} \rightarrow l^+l^-\,, \quad g+g \rightarrow l^+l^-\,.
\end{align}
\\
At next-to-leading order (NLO) in QCD, we have 

\begin{minipage}{2.5in}
\begin{align}
&q + \bar{q} \rightarrow l^+l^-+g\,,
\nonumber\\
&g + g \rightarrow l^+l^- + g\,,
\nonumber\\
&g + q \rightarrow l^+l^- + q\,,
\nonumber
\end{align}
\end{minipage}
\begin{minipage}{2.5in}
\begin{align}
\label{eq:17}
&q + \bar{q} \rightarrow l^+l^-+ \text{one loop}\,,
\nonumber\\
&g + g \rightarrow l^+l^-   + \text{one loop}\,,
\nonumber\\
&g  +\bar{q} \rightarrow l^+l^-  + \bar{q}\,.
\end{align}
\end{minipage}
\\

\noindent
At NNLO level, we have double real emission, 

\begin{minipage}{2.5in}
\begin{align*}
&q+\bar{q} \rightarrow l^+l^-  + q + \bar{q}\,,
\nonumber\\
&g+g \rightarrow l^+l^-  + g + g\,,
\nonumber\\
&g+q \rightarrow l^+l^-  + g + q\,,
\nonumber\\
&q+q \rightarrow l^+l^-  + q + q\,,
\nonumber\\
&q_{1}+\bar{q}_{2} \rightarrow l^+l^-  + q_{1} + \bar{q}_{2}\,,
\nonumber
\end{align*}
\end{minipage}
\begin{minipage}{3.1in}
\begin{align}
\label{eq:18}
&q_1+\bar{q}_1 \rightarrow l^+l^-  + q_2 + \bar{q}_2\,,
\nonumber\\
&q+\bar{q} \rightarrow l^+l^-  + g + g\,,
\nonumber\\
&g+g\rightarrow l^+l^-  + q + \bar{q}\,,
\nonumber\\
&g+\bar{q} \rightarrow l^+l^-  + g + \bar{q}\,,
\nonumber\\
&q_{1}+q_{2} \rightarrow l^+l^-  + q_{1} + q_{2}\,,
\end{align}
\end{minipage}
\\ 

\noindent
single real emission at one loop, 

\begin{minipage}{2.5in}
\begin{align*}
&q + \bar{q} \rightarrow l^+l^-  + g +\text{one loop}\,,
\nonumber\\
&g + q \rightarrow l^+l^-  + q + \text{one loop}\,,
\nonumber
\end{align*}  
\end{minipage}
\begin{minipage}{3.1in}
\begin{align}
\label{eq:19}
&g+g \rightarrow l^+l^-  + g + \text{one loop}\,,
\nonumber\\
&g + \bar{q} \rightarrow l^+l^-  + \bar{q} + \text{one loop} \,,
\end{align}  
\end{minipage}
\\

\noindent 
and the pure double virtual diagrams:

\begin{align}
\label{eq:20}
&q + \bar{q} \rightarrow l^+l^-  + \text{two loop}\,,
\nonumber\\
&g + g \rightarrow l^+l^-  + \text{two loop}\,.
\end{align}  
\\
The virtual corrections at one and two loop levels are straightforward for this process, the 
phase space integrals are often hard to evaluate.
In the first computation of the NNLO QCD correction to the DY pair production
~\cite{Hamberg:1990np}, the phase space integrals were performed
in three different frames to achieve the final result. 
This method was successfully applied  
in~\cite{Ravindran:2003um} to obtain inclusive cross section for the Higgs production at NNLO. 
In~\cite{Harlander:2002wh}, using a systematic expansion around threshold, all the phase space integrals were performed
to obtain the partonic cross sections for both DY and Higgs productions at NNLO level.  Later on, 
in~\cite{Anastasiou:2002yz}, an elegant formalism was developed to compute both 
real emissions as well as virtual corrections
applying integration by parts 
(IBP)~\cite{Tkachov:1981wb, Chetyrkin:1981qh} and
Lorentz invariance (LI)~\cite{Gehrmann:1999as} identities.  This approach is famously called 
the method of reverse unitarity. 
The resulting master integrals (MIs) were computed using the technique of differential equations.
The state-of-the-art result, namely, N$^3$LO QCD corrections to the inclusive Higgs boson
  production~\cite{Anastasiou:2014vaa, Anastasiou:2015ema, Anastasiou:2016cez} uses the method of reverse unitarity. 
  We have systematically used this approach 
~\cite{Anastasiou:2002yz} to calculate the partonic cross section of
the DY pair production 
through intermediate spin-2 particle at NNLO QCD.   

Ultraviolet (UV), soft and
collinear (IR) divergences do show up beyond leading order 
and they are regularised in dimensional regularisation where the space-time
dimensions $d$ is chosen to be equal to $4+\epsilon$. 
The soft divergences cancel among virtual and real subprocesses processes 
thanks to Kinoshita-Lee-Nauenberg (KLN) theorem~\cite{Kinoshita:1962ur, Lee:1964is} 
and the remaining UV divergences as well as
initial state collinear divergences are removed in ${\overline {\text{MS}}}$ scheme
using UV renormalisation constants and mass factorisation kernels denoted by $\Gamma_{ab}(\mu_F)$
respectively.  Here, $\mu_{F}$ is the factorisation scale.
For the UV renormalisation, we need to perform 
renormalisation for strong coupling constant $a_s = g_s^2/16 \pi^2$ through $Z_{a_{s}}$ as well as 
renormalisation of $\kappa_I$ through $Z_{IJ}$ listed in
the previous section.  For the former, we have
\begin{align}
\label{eq:21}
{\hat a}_{s} S_{\epsilon} = \left( \frac{\mu^{2}}{\mu_{R}^{2}}  \right)^{\epsilon/2}
  Z_{a_{s}} a_{s} \,,
\end{align}
where,
\begin{align}
\label{eq:22}
&Z_{a_{s}} = 1+ a_s\left[\frac{2}{\epsilon} \beta_0\right]
             + a_s^2 \left[\frac{4}{\epsilon^2 } \beta_0^2
             + \frac{1}{\epsilon}  \beta_1 \right] + \cdot \cdot \cdot \,,
\end{align}  
$a_{s} \equiv a_{s}(\mu_{R}^{2})$,
$S_{\epsilon} = {\rm exp} \left[ (\gamma_{E} - \ln 4\pi)\epsilon/2
\right]\,, \gamma_{E} = 0.5772\ldots\,,$ 
and the scale $\mu$ is introduced to keep the unrenormalised strong
coupling constant ${\hat a}_{s}$ dimensionless in $n$-dimensions. The
renormalisation scale is denoted by $\mu_{R}$. $\beta_{i}$'s are the coefficients of QCD
$\beta$-function~\cite{Gross:1973id, Politzer:1973fx, Caswell:1974gg, Tarasov:2013zv, Larin:1993tp}. 
The mass factorised finite cross section can be obtained using  
\begin{align}
\label{eq:23}
	2 s {d \hat \sigma_{ab} \over dQ^2}(z,Q^2,1/\epsilon) =
  \sum_{c,d=q,{\bar q}, g}\Gamma_{ca}(z,\mu_F^2,1/\epsilon) \otimes
  \Gamma_{db}(z,\mu_F^2,1/\epsilon)\otimes 
   2  s {d \sigma_{ab} \over dQ^2}(z,Q^2,\mu_{F}^2)\,,
\end{align}
where $\otimes$ are nothing but Mellin convolution.  The mass factorisation kernels take the following form
\begin{align}
\label{eq:24}
	\Gamma_{ab}(z, \mu_{F}^{2},1/\epsilon) =& 
\delta_{ab} \delta(1-z)
        + a_s(\mu_F^2) \frac{1}{\epsilon}  P^{(0)}_{ab}(z)
\nonumber \\
	& +a_{s}^2(\mu_{F}^{2})
	\left[\frac{1}{\epsilon^{2}} \Bigg( \frac{1}{2} P^{(0)}_{ac} \otimes
  P^{(0)}_{cb} + \beta_{0} P^{(0)}_{ab}\Bigg) + \frac{1}{\epsilon}
	\Bigg( \frac{1}{2} P^{(1)}_{ab}\Bigg)\right] + \cdot \cdot \cdot \,, 
\end{align}
where $P^{(i)}_{ab}$ are the Altarelli-Parisi splitting
functions~\cite{Altarelli:1977zs, Floratos:1980hm, Floratos:1980hk, Curci:1980uw, Moch:2004pa,Vogt:2004mw}.
After the mass factorisation, the finite partonic cross sections denoted by  $2 s {d \sigma_{ab}/ dQ^2}$ can be expressed in terms
$\Delta^h_{ab}(z,a_s(\mu_R^2),Q^2/\mu_R^2,\mu_F^2/\mu_R^2)$ by factoring out some overall constants. In terms of these $\Delta^h_{ab}$, the 
hadronic cross section can be written as
\begin{align}
	2 S{d \sigma^{H_1H_2}_{\rm spin-2} \over dQ^2}(\tau,Q^2)&=
	\sum_{q,\bar q,g}{\cal F}_{h} \int_0^1 {d x_1 } \int_0^1 
{dx_2} \int_0^1 dz \delta(\tau-z x_1 x_2)
\times \Bigg[ 
H_{q{\bar q}} 
             \sum\limits_{k=0}^{2} a_{s}^{k} \Delta^{h, (k)}_{q{\bar q}} 
\nonumber\\&
+
H_{g g} \sum\limits_{k=0}^{2} a_{s}^{k} \Delta^{h, (k)}_{gg} 
+ \Big( H_{gq} + H_{qg}  \Big)
             \sum\limits_{k=1}^{2} a_{s}^{k} \Delta^{h, (k)}_{gq}
\nonumber\\&+
H_{q q} \sum\limits_{k=2}^{2}
             a_{s}^{k} \Delta^{h, (k)}_{qq}  
+ H_{q_{1} q_{2}}  \sum\limits_{k=2}^{2}
             a_{s}^{k} \Delta^{h, (k)}_{q_{1}q_{2}}  
\Bigg]\,,
\end{align}
where
\begin{align}
\label{eq:31}
{\cal F}_{h}=\;{\kappa_Q^2 Q^6 \over 320 \pi^2 }|{\cal D}(Q^2)|^2\,,
\quad \quad \quad
	\Delta^{h,(k)}_{ab} ~= \Delta^{h,(k)}_{ab} \left(z,{Q^2\over \mu_R^2},{\mu_F^2\over \mu_R^2}\right) \,.
\end{align}
$\kappa_Q$ in ${\cal F}_{h}$ corresponds to the leptonic coupling to the spin-2, while
the coupling to quarks and gluons are taken in $\Delta^{h,(k)}_{ab}$. We have provided analytical expressions for these  $\Delta^{h,(k)}_{ab}$ in Mathematica format as an ancillary file.
${\cal D}(Q^2)$ is the propagator of the massive spin-2 particle, with a decay width that has
to be estimated considering its decay to SM particles.
$H_{ab}$ are the combinations of the mass factorised partonic distribution functions:  
\begin{align}
\label{eq:32}
H_{q \bar q}(x_1,x_2,\mu_F^2)&=
f_q^{H_1}(x_1,\mu_F^2) 
f_{\bar q}^{H_2}(x_2,\mu_F^2)
+f_{\bar q}^{H_1}(x_1,\mu_F^2)~ 
f_q^{H_2}(x_2,\mu_F^2)\,,
\nonumber\\
H_{q q}(x_1,x_2,\mu_F^2)&=
f_q^{H_1}(x_1,\mu_F^2) 
f_{q}^{H_2}(x_2,\mu_F^2)
+f_{\bar q}^{H_1}(x_1,\mu_F^2)~ 
f_{\bar q}^{H_2}(x_2,\mu_F^2)\,,
\nonumber\\
H_{q_1 q_2}(x_1,x_2,\mu_F^2)&=
f_{q_1}^{H_1}(x_1,\mu_F^2) 
\Big( f_{q_2}^{H_2}(x_2,\mu_F^2) + f_{\bar q_2}^{H_2}(x_2,\mu_F^2) \Big)
\nonumber\\&+f_{\bar q_1}^{H_1}(x_1,\mu_F^2)~ 
\Big( f_{q_2}^{H_2}(x_2,\mu_F^2) + f_{\bar q_2}^{H_2}(x_2,\mu_F^2) \Big)\,,
\nonumber\\
H_{g q}(x_1,x_2,\mu_F^2)&=
f_g^{H_1}(x_1,\mu_F^2) 
\Big(f_q^{H_2}(x_2,\mu_F^2)
+f_{\bar q}^{H_2}(x_2,\mu_F^2)\Big)\,,
\nonumber\\
H_{q g}(x_1,x_2,\mu_F^2)&=
H_{g q}(x_2,x_1,\mu_F^2)\,,
\nonumber\\
H_{g g}(x_1,x_2,\mu_F^2)&=
f_g^{H_1}(x_1,\mu_F^2)~ 
f_g^{H_2}(x_2,\mu_F^2)\,.
\end{align}
In the next section, we study the numerical implication of NNLO QCD corrections 
to a spin-2 coupling non-universally to the SM in the DY process.

\begin{section}{Numerical results}
In this section, we present the numerical impact of our NNLO results on the
production of di-leptons at the LHC.  We considered a minimal scenario of
non-universal couplings
of spin-2 particle with SM fields, where the spin-2 particle couples to all SM fermions 
with coupling $\kappa_Q = \sqrt{2} k_q/\Lambda$ and  to all SM gauge bosons with a
coupling strength of $\kappa_G = \sqrt{2} k_g/\Lambda$.  Numerical results presented
in this section are for the default choice of model parameters, namely spin-2 particle
of mass $m_G=500$ GeV, the scale $\Lambda=2$ TeV and the couplings $(k_q,k_g) = (0.5,1.0)$. Both the renormalization and factorization scales are set equal to the invariant mass of the di-lepton, i.e.,
$\mu_R = \mu_F = Q$. Throughout, we use MSTW2008nnlo parton distribution functions (PDFs) with the corresponding $a_s$ 
	provided from {\tt LHAPDF} unless otherwise stated.  The choose  $\sqrt{S}=13$ TeV, the center of mass
energy of the incoming hadrons at the LHC.

In our analysis, we restricted ourselves to the situation where spin-2 particle
decays only to SM fields. The spin-2 particle decay widths for non-universal
couplings are same as those given in \cite{Han:1998sg}. For the scenario taken
up here, where in all spin-2 coupling to all bosons are taken to be identical,
we note that
spin-2 particle decaying to $Z \gamma$ vanishes identically 
$\Gamma(h \to Z\gamma) =0$ \cite{Falkowski:2016glr}.  In fig.\ref{subnlo}, we
present the NLO corrections
(only at order $a_s$) from various subprocess contributions to the di-lepton
production. For our default choice of model parameters, we find that $gg$ subprocess 
contribution dominates over the rest. In general, the total NLO correction is
smaller than the $gg$ contribution because of negative contribution from $qg$
subprocess. We also note that the $gg$ has dominant contribution to the total 
decay width for couplings (0.5, 1.0).

To estimate the impact of QCD corrections, we define the K-factors as follows:
\begin{eqnarray}
\text{K}_1 = \frac{d\sigma^{\text{NLO}}/dQ}{d\sigma^{\text{LO}}/dQ}  
\qquad \text{and} \qquad 
\text{K}_2 = \frac{d\sigma^{\text{NNLO}}/dQ}{d\sigma^{\text{LO}}/dQ}.
\label{eqnkf}
\end{eqnarray}
In the left panel of fig.\ref{nlokf}, we present di-lepton invariant mass distributions to NLO for different choices of non-universal 
couplings $(k_q,k_g)= (1.0,0.5), (1.0,0.1)$ and $(0.5,0.1)$. It is expected for universal couplings that at the resonance region,
the cross sections i.e. the height of the peak will be the same simply because the couplings at the matrix element level will cancel with those
from the decay width of the spin-2 particle. However, for non-universal couplings this is not the case 
and hence cross sections at the resonance for different non-universal couplings will be different.
Thus, the precision as well as the phenomenological studies of the spin-2 particle production in this model
will be different from those of the warped extra dimension models.The NLO K-factor ($K_1$) is present in the right panel for various choices of $(k_q,k_g)$ and we observe that the K-factor crucially depends on the choice
of non-universal couplings. In particular we notice that the K-factors are larger for the choice of couplings (1.0,0.1). To understand
this behaviour better, it is helpful to study the percentage contribution of various subprocesses to the total correction
at NLO level, particularly  from $qg$ subprocess due to its large flux at LHC energies. In particular we define the percentage of contribution 
of a given subprocess $ab$ as $R^{(i)}_{ab} = (d\sigma^{H_1 H_2,(i)}_{ab}/dQ^2) /
	(d \sigma^{H_1 H_2,(i)}/dQ^2) \times 100$, where the numerator is obtained by using 
contribution from $\Delta^{h,(i)}_{ab}$ and for the denominator, we include all the partonic channels.

In fig.\ref{qgnlo-fr}, we plot $R^{(1)}_{qg}$ for different choices of non-universal couplings and we observe that the 
sign of the $qg$ subprocess crucially depends on the choice of couplings. Moreover, we find that $R^{(1)}_{qg}$ is positive
and is as large as $70\%$ for the couplings $(1.0,0.1)$, which explains the reason for the large K-factor at the resonance
region. However, the sign of the contribution from other subprocesses $q\bar{q}$ and $gg$ is found 
to be positive for various couplings.

\begin{figure}[htb]
\centerline{
\epsfig{file=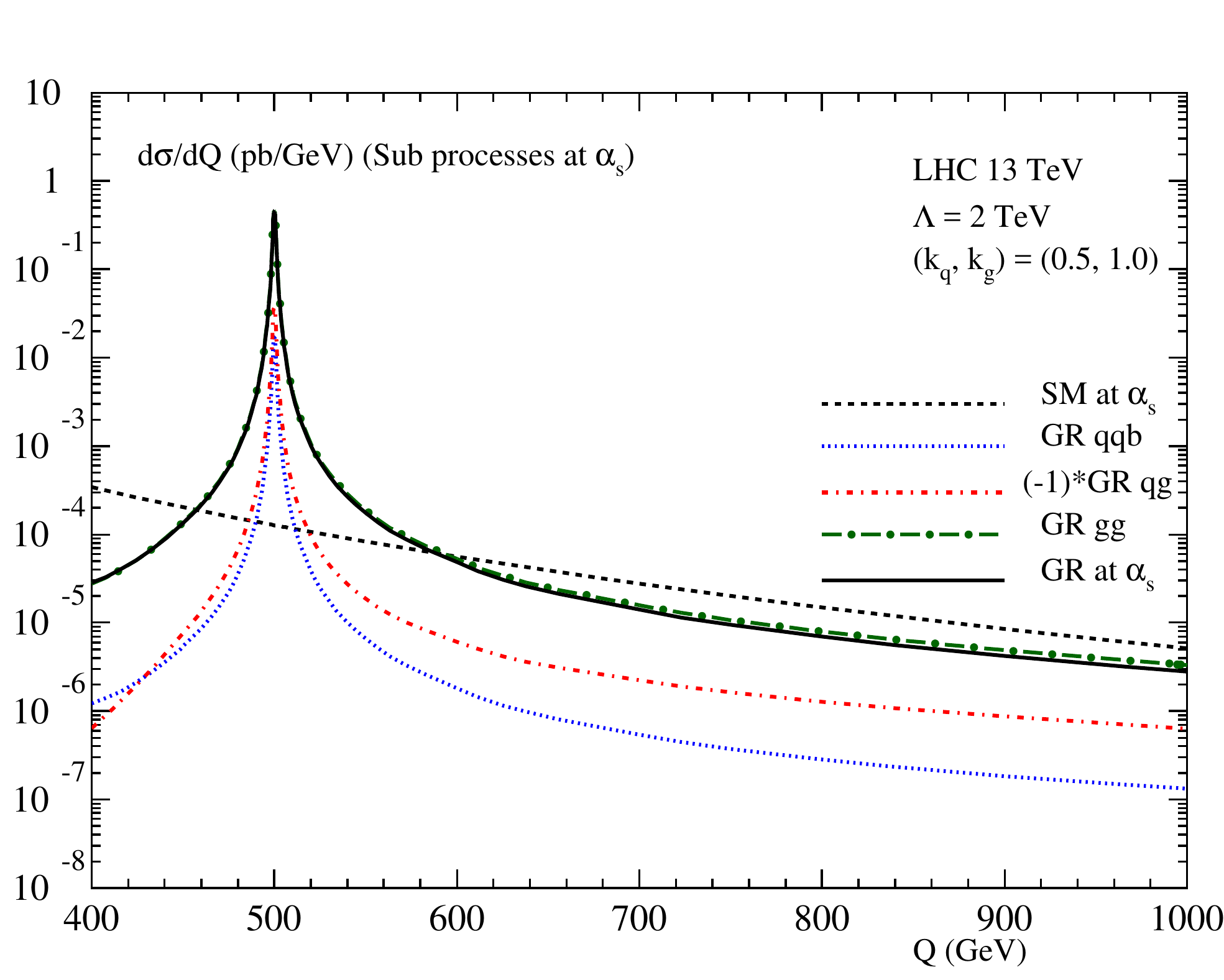,width=8.0cm,height=8.0cm,angle=0}
}
\caption{\sf {First order QCD corrections from different subprocesses to di-lepton production. The choice of the
model parameters is as mentioned in the text.}}
\label{subnlo}
\end{figure}

\begin{figure}[htb]
\centerline{
\epsfig{file=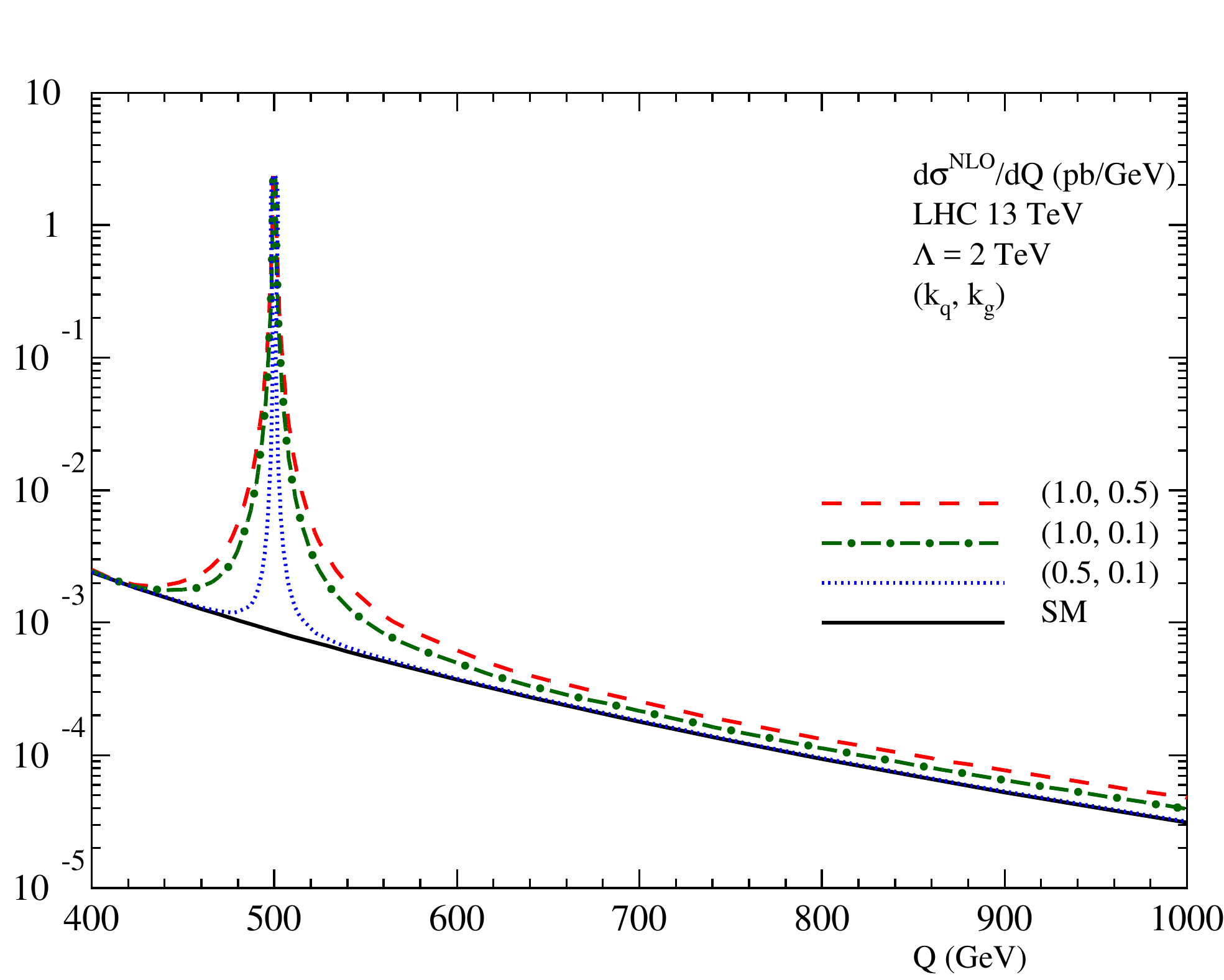,width=7.5cm,height=7.5cm,angle=0}
\epsfig{file=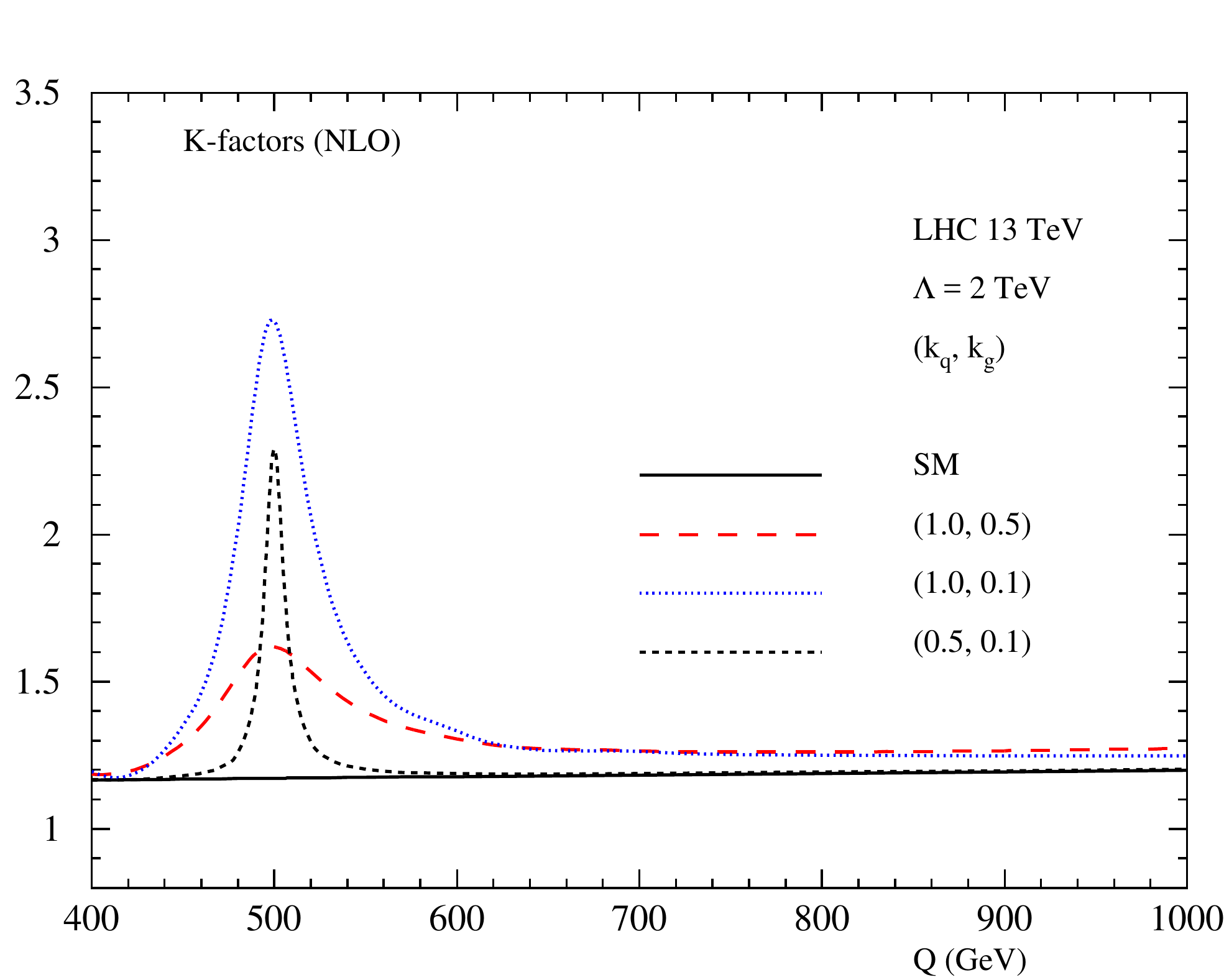,width=7.5cm,height=7.5cm,angle=0}
}
\caption{\sf {Di-lepton invariant mass distributions are presented to NLO QCD for
different choice of couplings ($k_q, k_g$)} in the left panel. The corresponding K-factors
are presented in the right panel.}
\label{nlokf}
\end{figure}

\begin{figure}[htb]
\centerline{
\epsfig{file=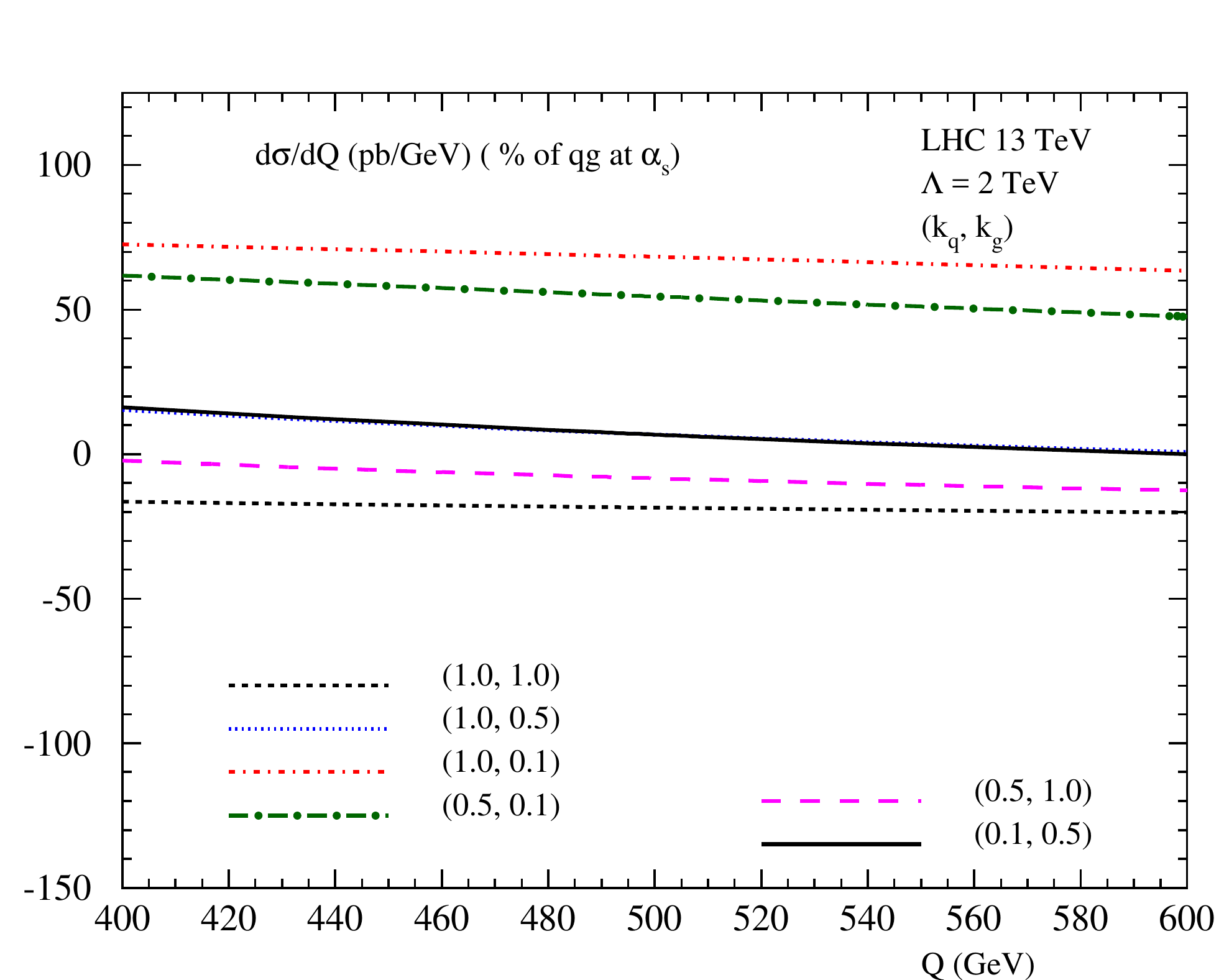,width=7.50cm,height=7.50cm,angle=0}
}
\caption{\sf {Percentage of $qg$ subprocess contribution $R^{(1)}_{qg}$ as defined in the text for different
choice of non-universal couplings.}}
\label{qgnlo-fr}
\end{figure}

In fig.\ref{subnnlo},  we present the second order QCD corrections (at $(a_s^2)$) 
from various subprocesses to the di-lepton production for the default choice of couplings
$(k_q, k_g) = (0.5,1.0)$.  Similar to the first order QCD corrections, $gg$ 
subprocess has the dominant contribution over the rest while $qg$ has a negative contribution
but is comparable in magnitude to that of $gg$. Because of this large $qg$ subprocess 
contribution which can flip its sign for certain couplings, it is necessary to study 
the percentage of its relative contribution $R^{(2)}_{qg}$ to the total second order 
correction.
\begin{figure}[htb]
\centerline{
\epsfig{file=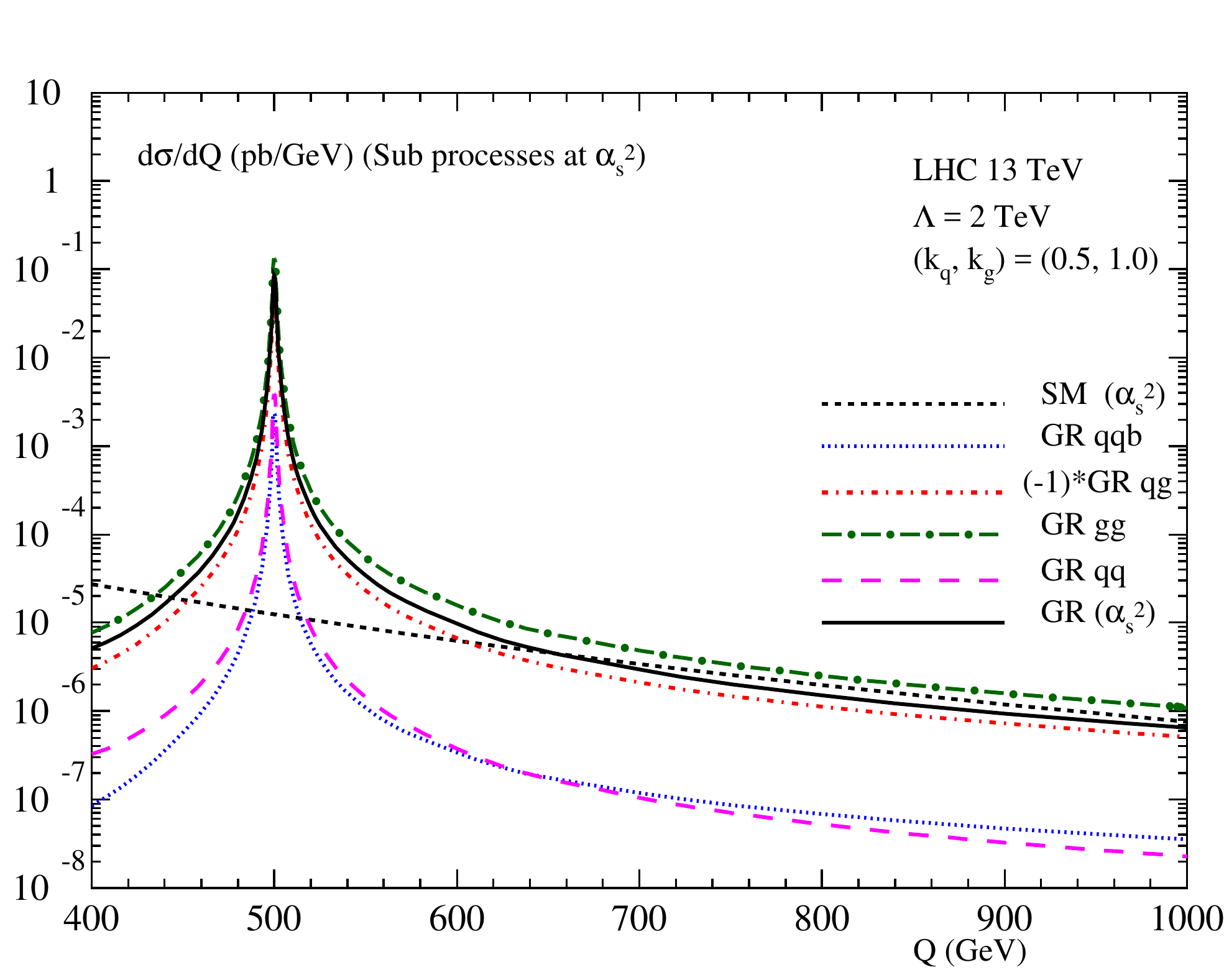,width=7.50cm,height=7.50cm,angle=0}
}
\caption{\sf {Second order QCD corrections from various subprocess to the di-lepton invariant
mass distribution.}}
\label{subnnlo}
\end{figure}
In fig.\ref{fr-qgnnlo}, we present $R^{(2)}_{qg}$ for different
choices of couplings. As can be seen from the figure,  the $qg$ contribution varies
from about $-70\%$ to about $35\%$ for the choice of couplings considered here.
In particular, for the couplings $(1.0,0.1)$ and $(0.5, 0.1)$ the $qg$ contribution
is positive while it is negative for the rest of the couplings as well as in the SM.
This implies large K-factors for the choice of $(1.0, 0.1)$ couplings for a wide
range of the invariant mass distribution.
It is worth mentioning here that in general $qg$ subprocess has a negative contribution
both in the SM as well as in the case of universal couplings, irrespective of the value
of the latter.

We then present the di-lepton invariant mass distribution to various orders in QCD
for a particular choice of couplings $(1.0, 0.5)$ in fig.\ref{c1order}. In this
case, the NLO QCD corrections for the signal (SM+spin-2) are as large as $60\%$ while those
at NNLO, they are about $80\%$ at the resonance. Similar results are presented but for our 
default choice of model parameters in fig.\ref{c2order}. Here, the corresponding NLO
corrections to the signal are about $45\%$ while those of NNLO are about $55\%$.
\begin{figure}[htb]
\centerline{
\epsfig{file=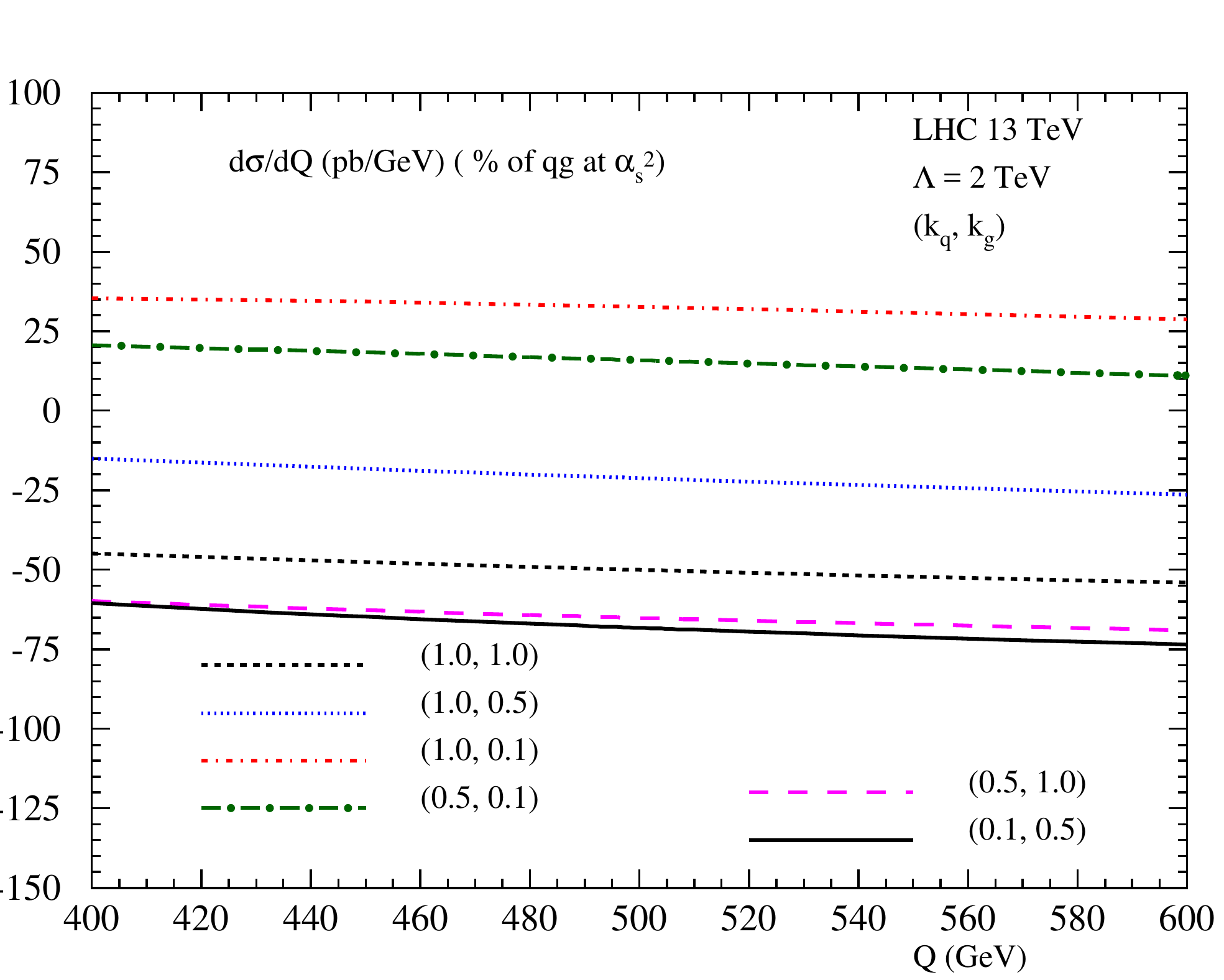,width=7.50cm,height=7.50cm,angle=0}
}
\caption{\sf {Percentage of $qg$ contribution $R_{qg}^{(2)}$} as defined in the text.}
\label{fr-qgnnlo}
\end{figure}

\begin{figure}[htb]
\centerline{
\epsfig{file=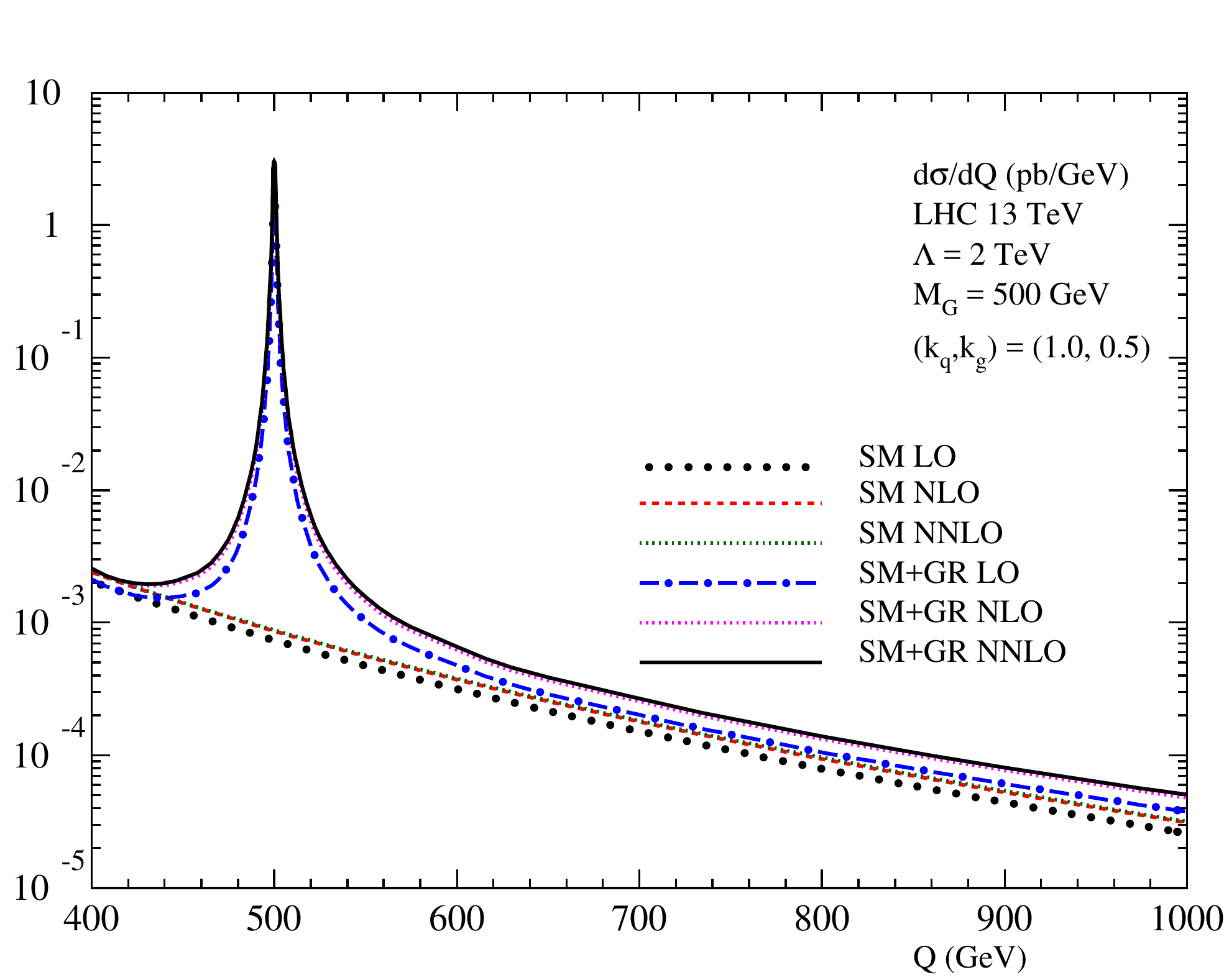,width=7.5cm,height=7.5cm,angle=0}
\epsfig{file=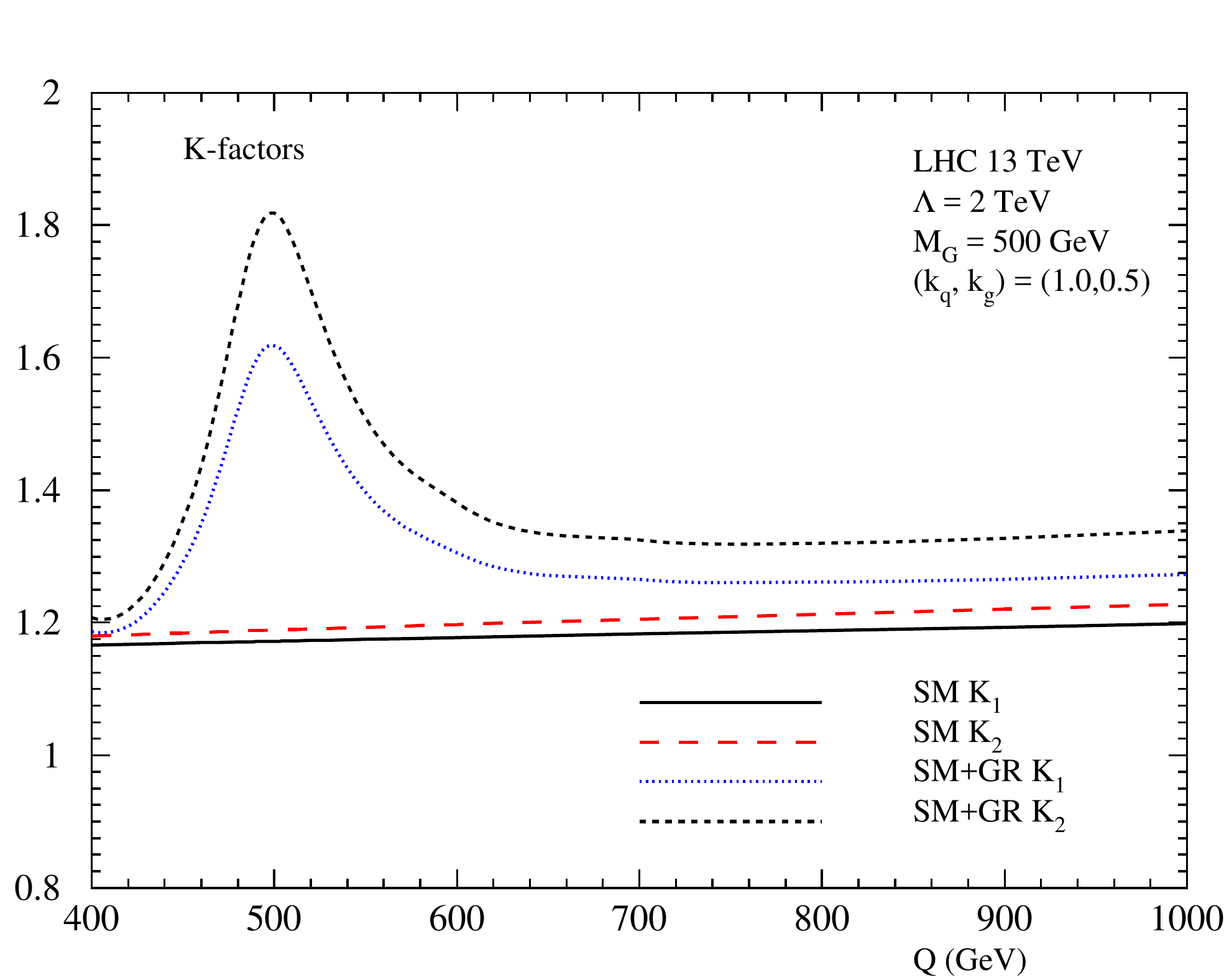,width=7.5cm,height=7.5cm,angle=0}
}
\caption{\sf {cross sections at different orders (left panel) and the corresponding K-factors $K_1$ 
and $K_2$ (right panel) are presented for different couplings.}}
\label{c1order}
\end{figure}

\begin{figure}[htb]
\centerline{
\epsfig{file=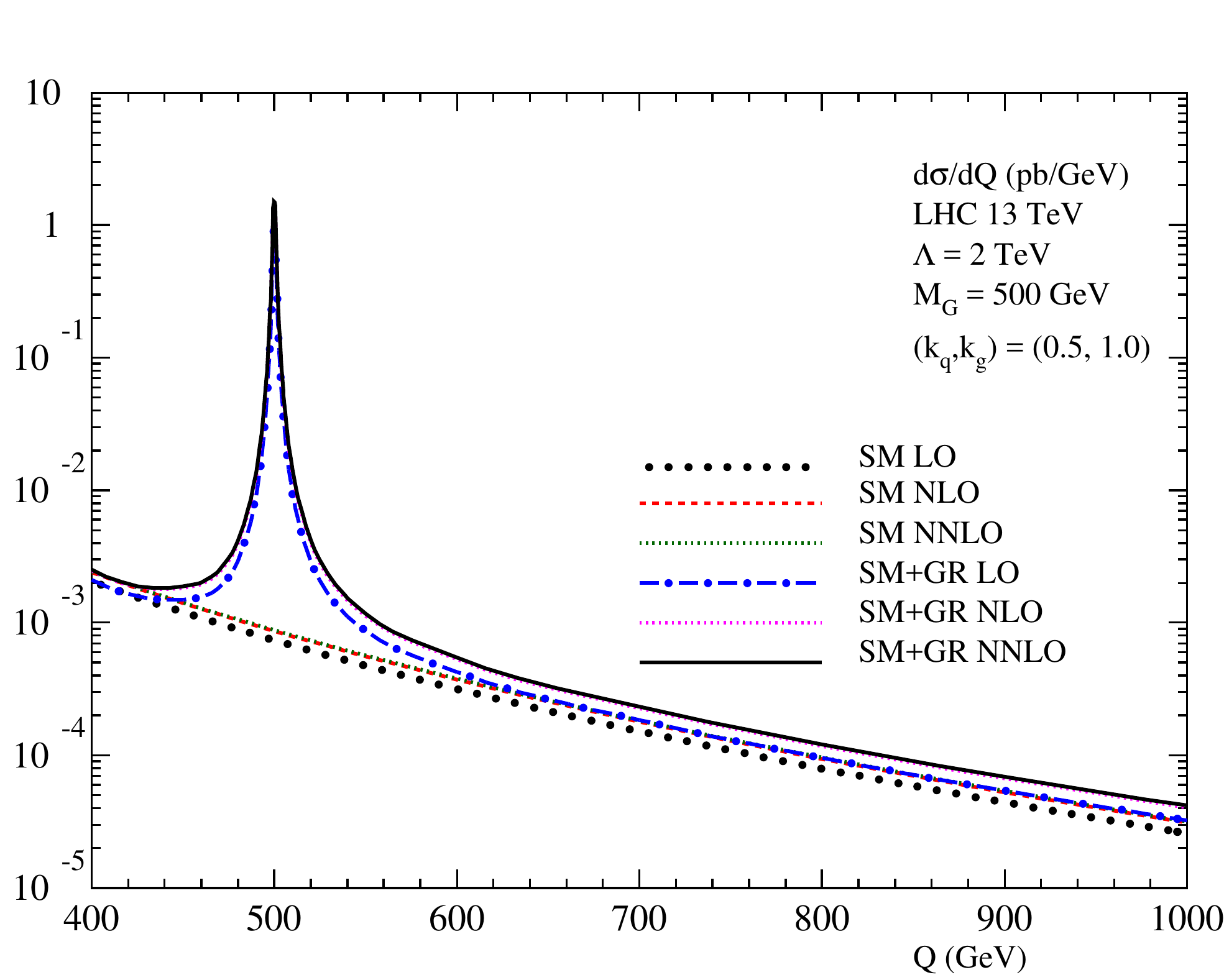, width=7.5cm,height=7.5cm,angle=0}
\epsfig{file=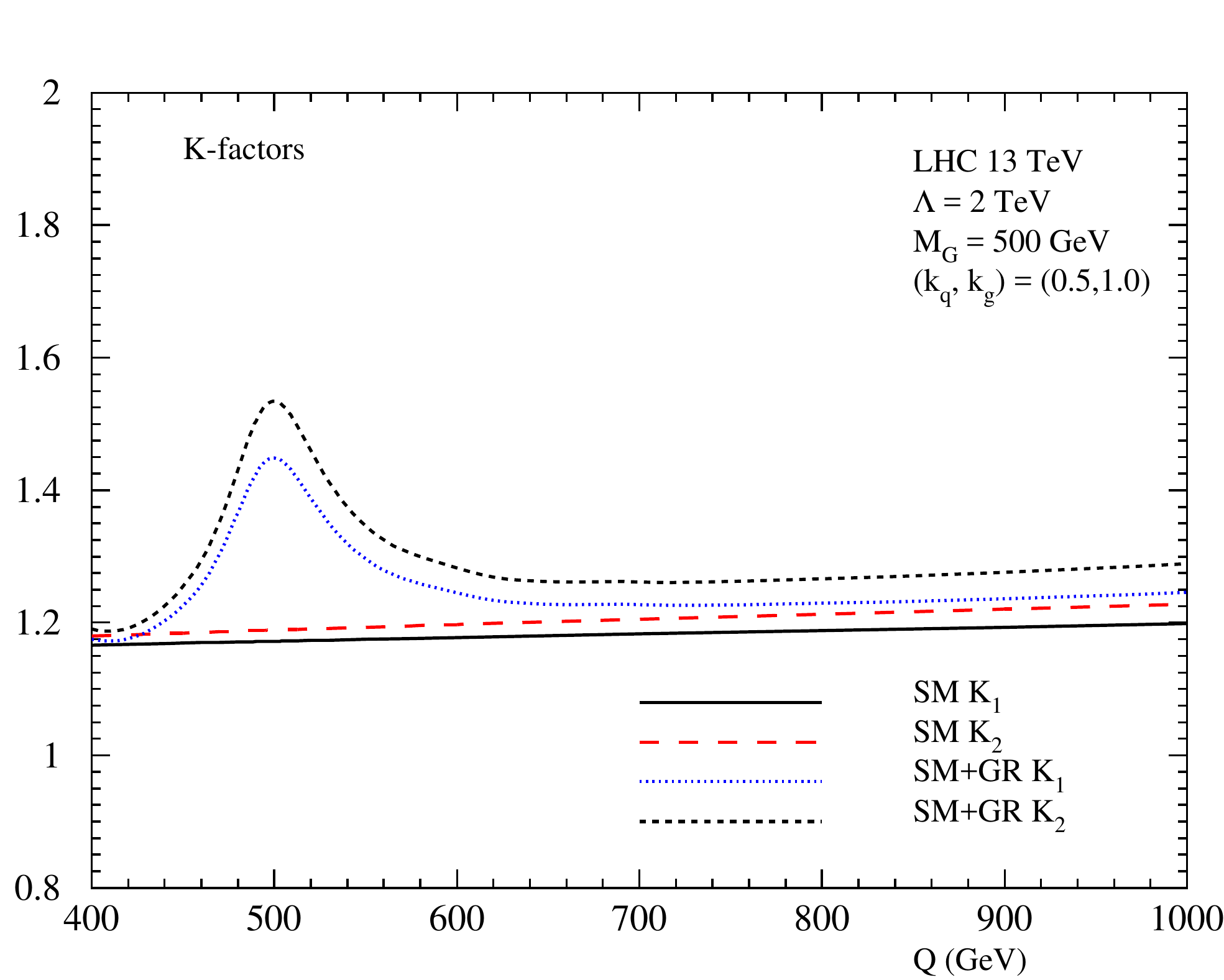, width=7.5cm,height=7.5cm,angle=0}
}
\caption{\sf {Same as fig.\ref{c1order} but for a different set of couplings.}}
\label{c2order}
\end{figure}

Next, we will study the invariant mass distributions of both the
SM and the signal, in particular the impact of QCD corrections for
different couplings. In fig.\ref{c01var},\ref{c02var},\ref{c03var}, we 
present these distributions in the left panel and the corresponding NNLO K-factors 
$(K_2)$ in the right panel for 9 different set of non-universal couplings.
The respective K-factors for the signal at the resonance region are found
to vary from about $1.5$ to about as large as $3.0$, owing to different
contributions from $qg$ subprocess to the signal as explained before.

\begin{figure}[htb]
\centerline{
\epsfig{file=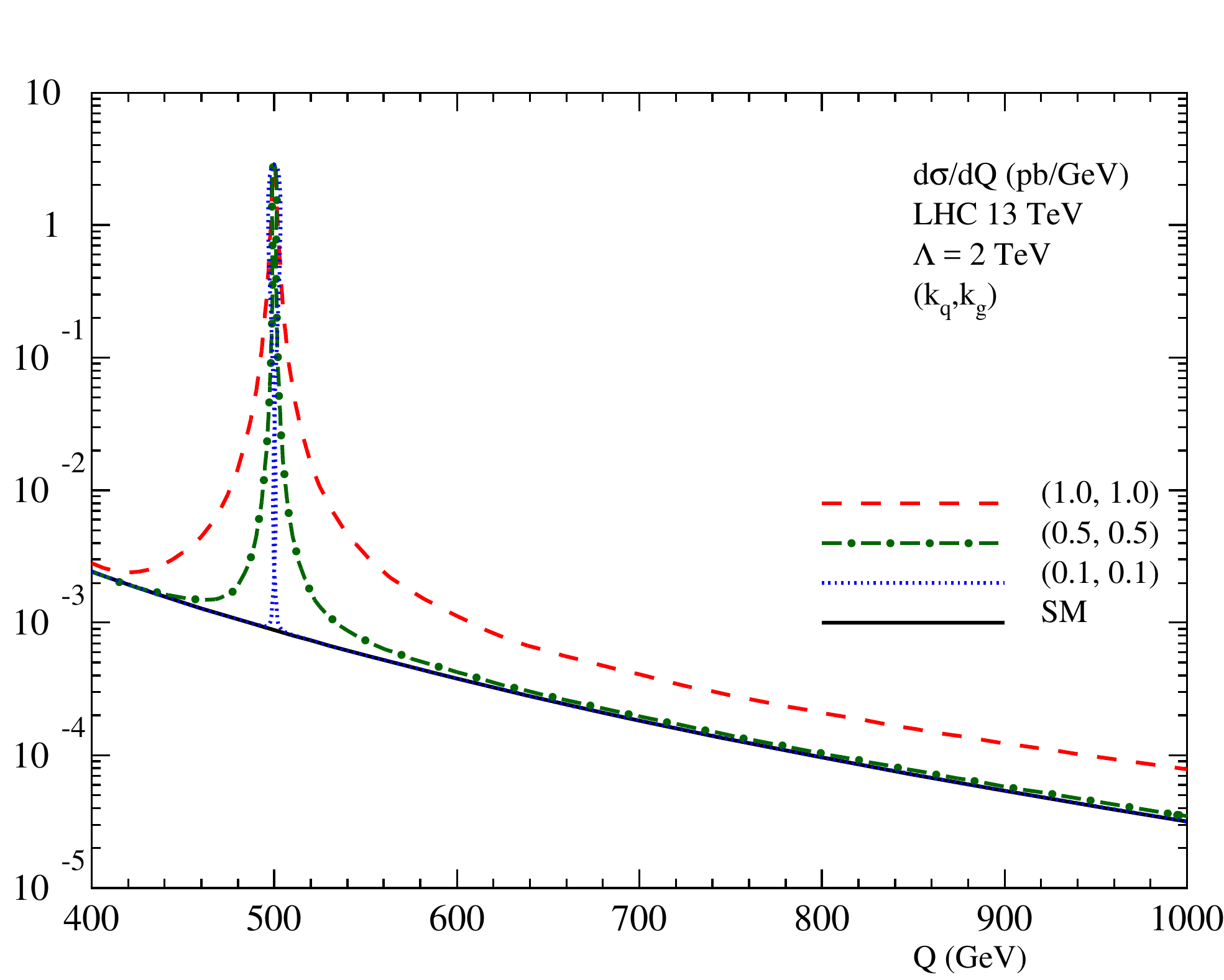,width=7.5cm,height=7.5cm,angle=0}
\epsfig{file=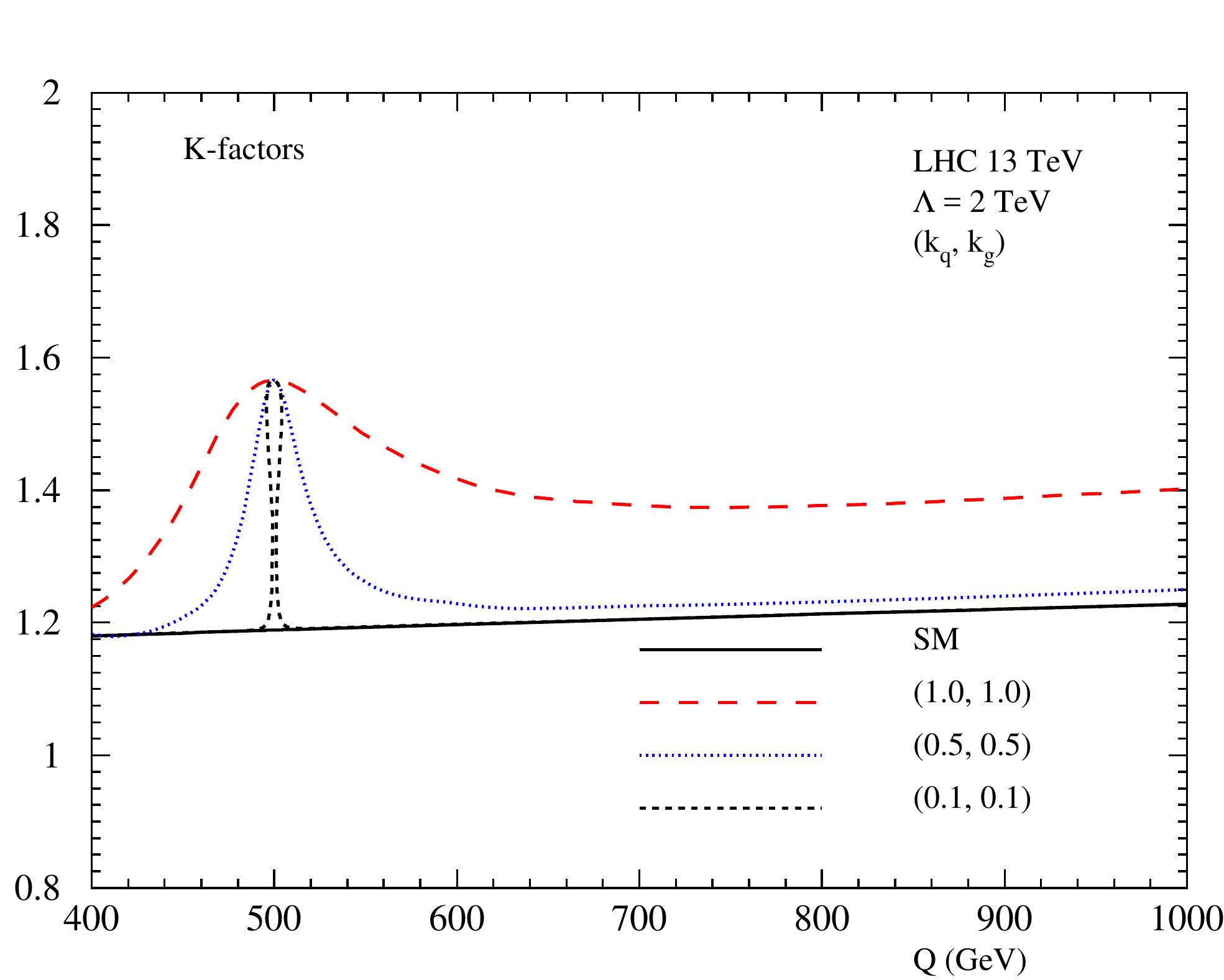,width=7.5cm,height=7.5cm,angle=0}
}
\caption{\sf {Di-lepton invariant mass distributions to NNLO for different choice
of couplings (left panel) and the corresponding K-factors (right panel) are presented.}}
\label{c01var}
\end{figure}

\begin{figure}[htb]
\centerline{
\epsfig{file=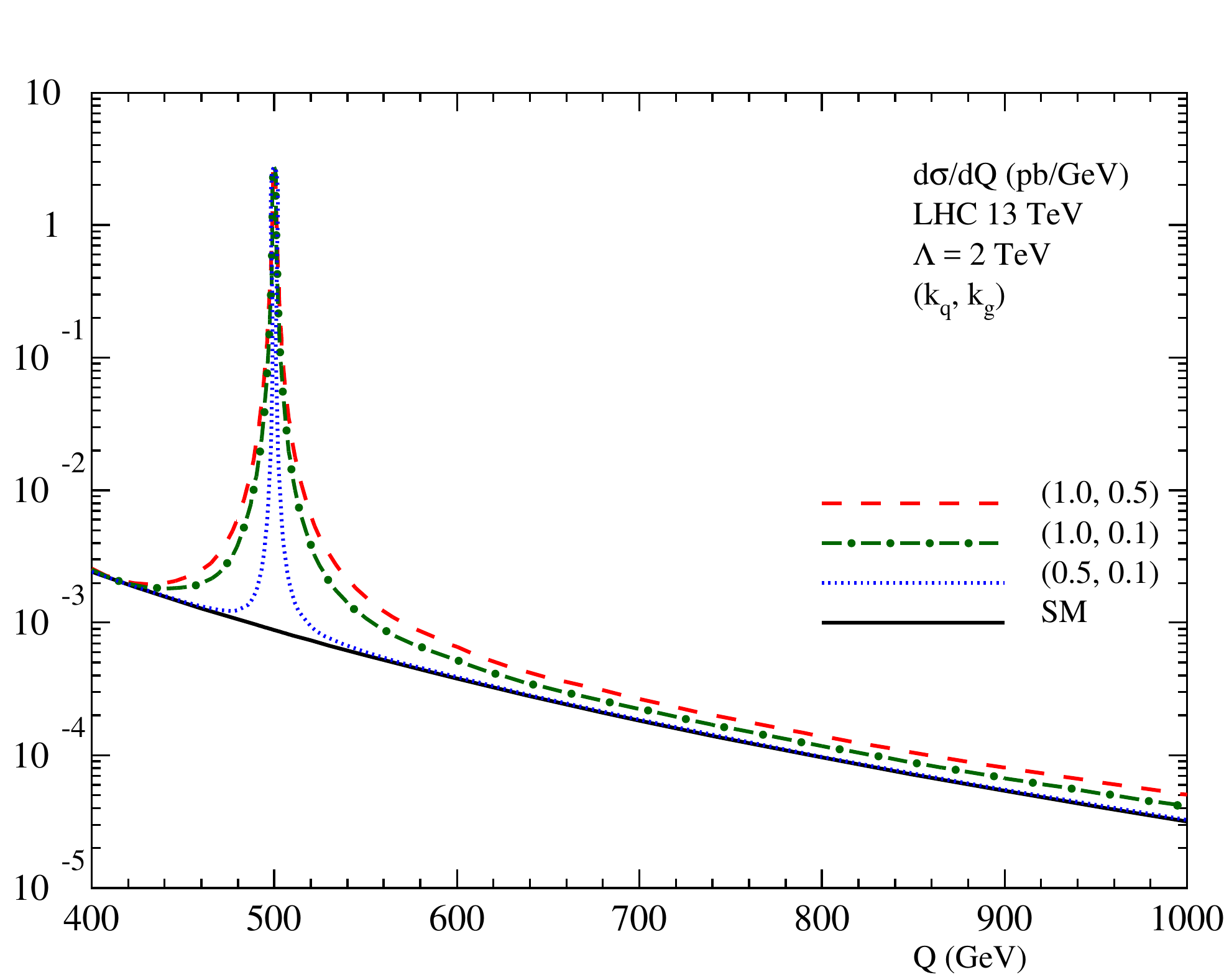,width=7.5cm,height=7.5cm,angle=0}
\epsfig{file=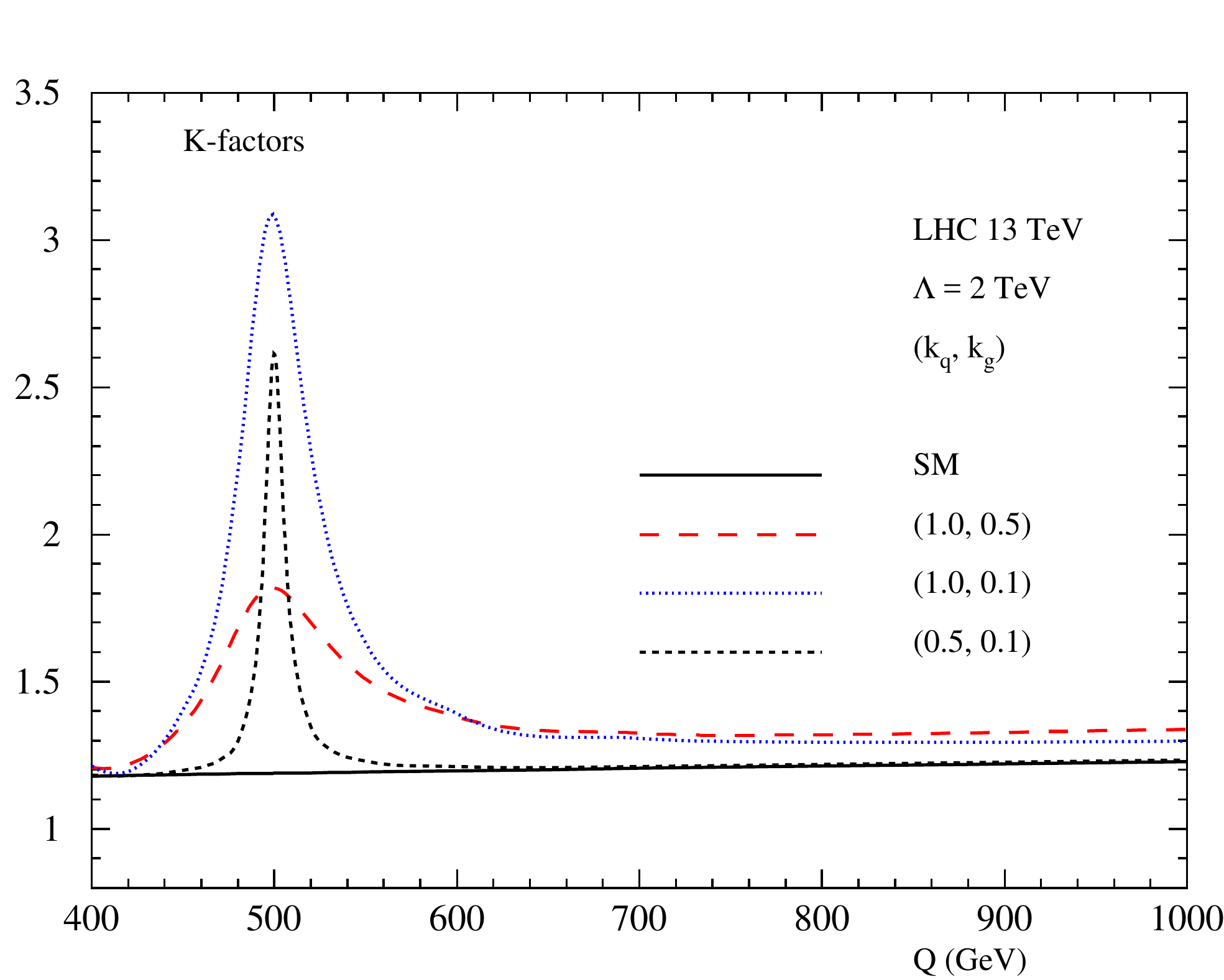,width=7.5cm,height=7.5cm,angle=0}
}
\caption{\sf {Same as fig.\ref{c01var} but for a different set of couplings.}}
\label{c02var}
\end{figure}

\begin{figure}[htb]
\centerline{
\epsfig{file=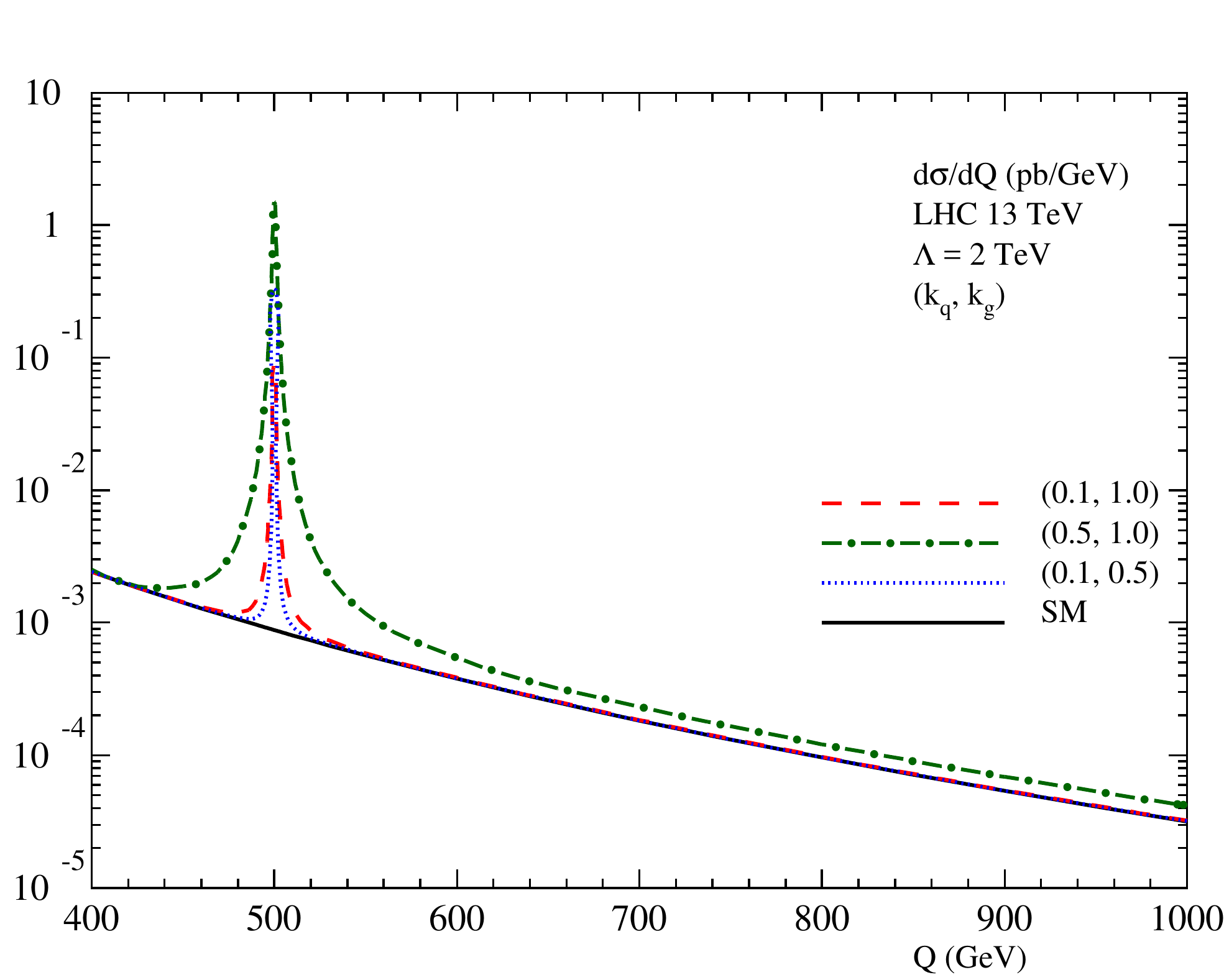,width=7.5cm,height=7.5cm,angle=0}
\epsfig{file=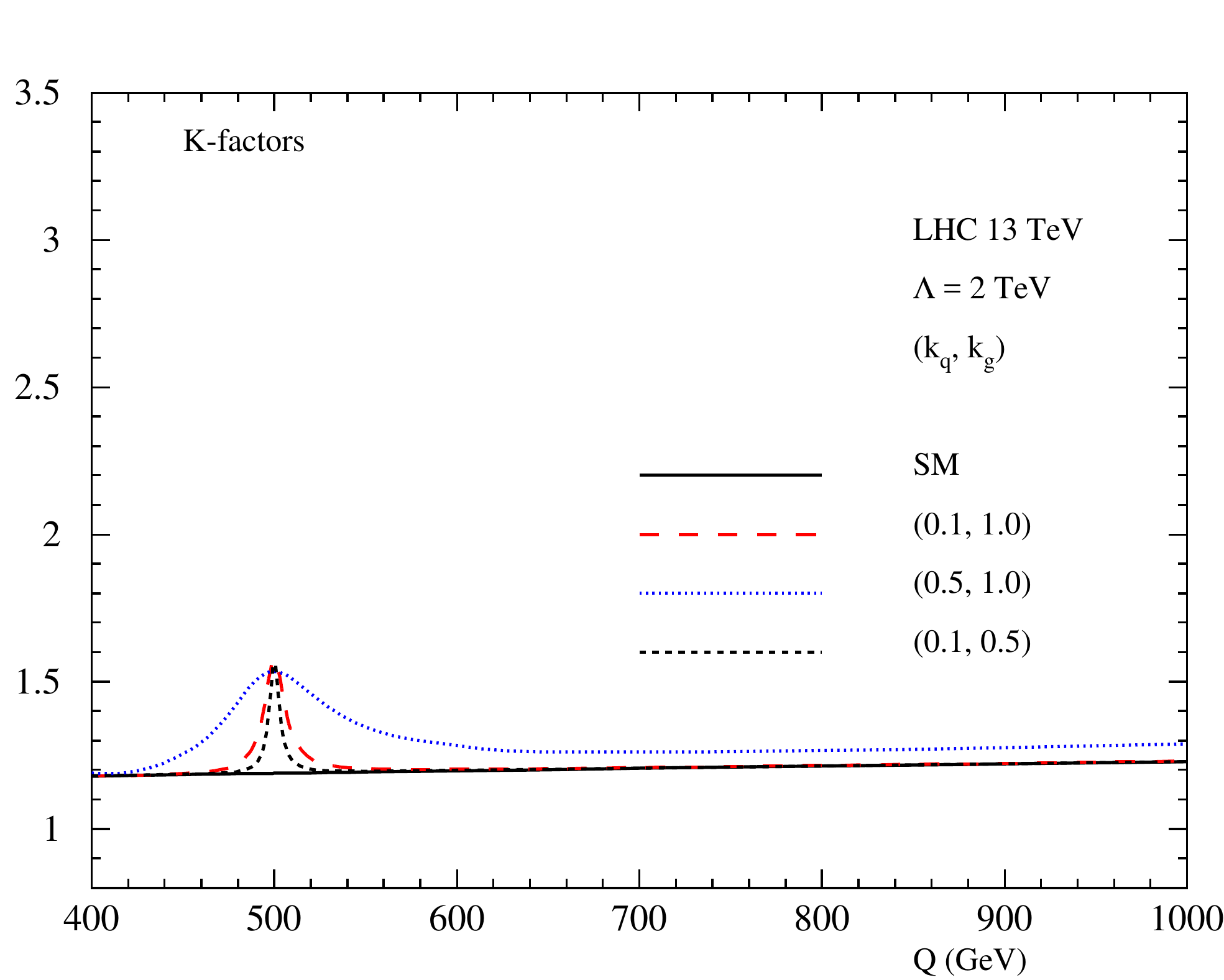,width=7.5cm,height=7.5cm,angle=0}
}
\caption{\sf {Same as fig.\ref{c01var} but for a different set of couplings.}}
\label{c03var}
\end{figure}

Further, we depict the dependence of invariant mass distributions to NNLO
in QCD on the center of mass energy E$_\text{cm}$ of the protons at the LHC.
We present our results for E$_\text{cm} = 7, 8, 13$ and $14$ TeV energies
for two different sets of couplings. In fig.\ref{ecm-var1}, we present
the invariant mass distributions and the corresponding K-factors for
the universal couplings of $(1.0, 1.0)$. For default choice of non-universal
couplings $(0.5, 1.0)$, similar results are presented in fig.\ref{ecm-var2}.
In both the cases, the K-factors at the resonance region 
are found to be larger for $7$ TeV case and are about $1.6$.

\begin{figure}[htb]
\centerline{
\epsfig{file=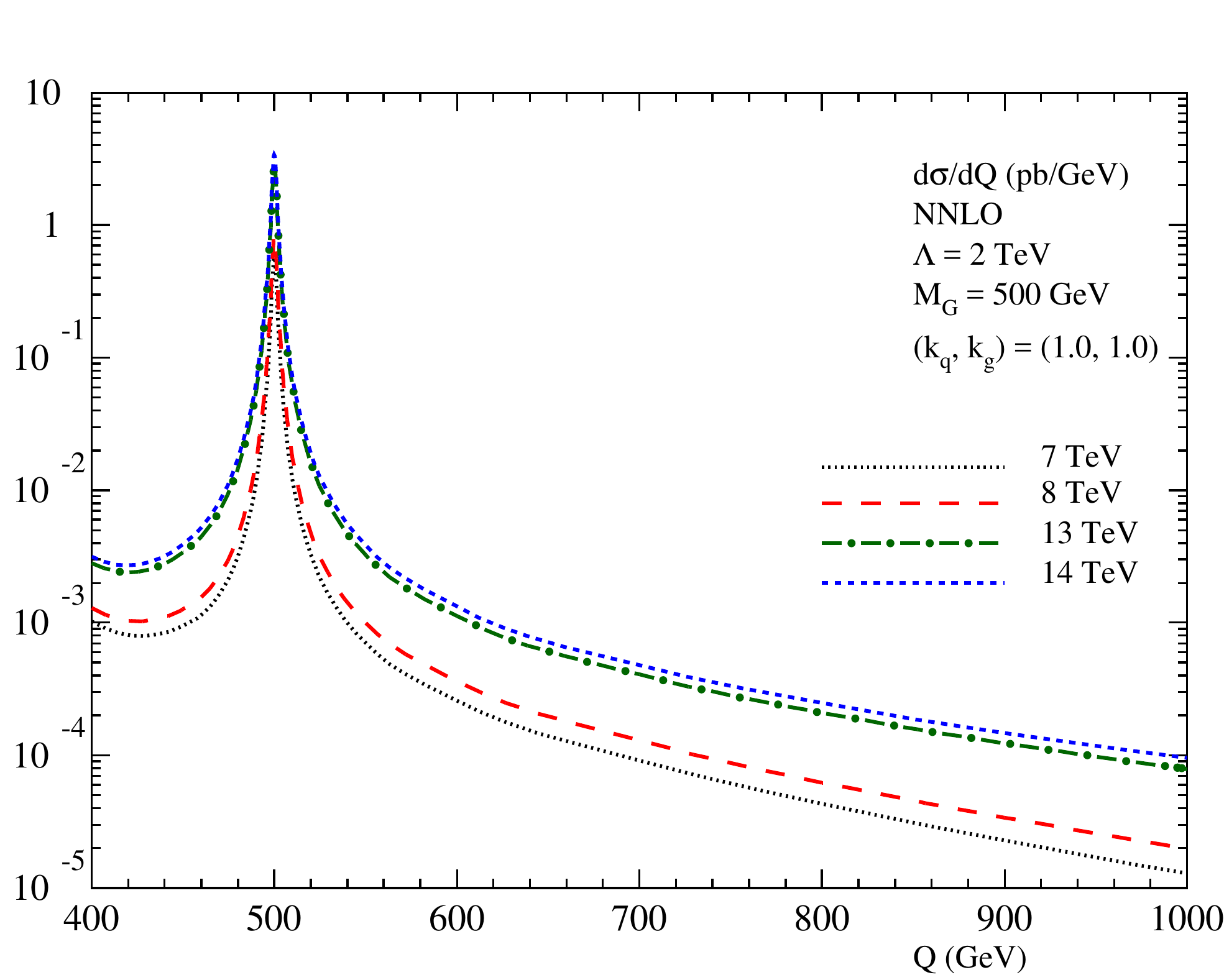,width=7.5cm,height=7.5cm,angle=0}
\epsfig{file=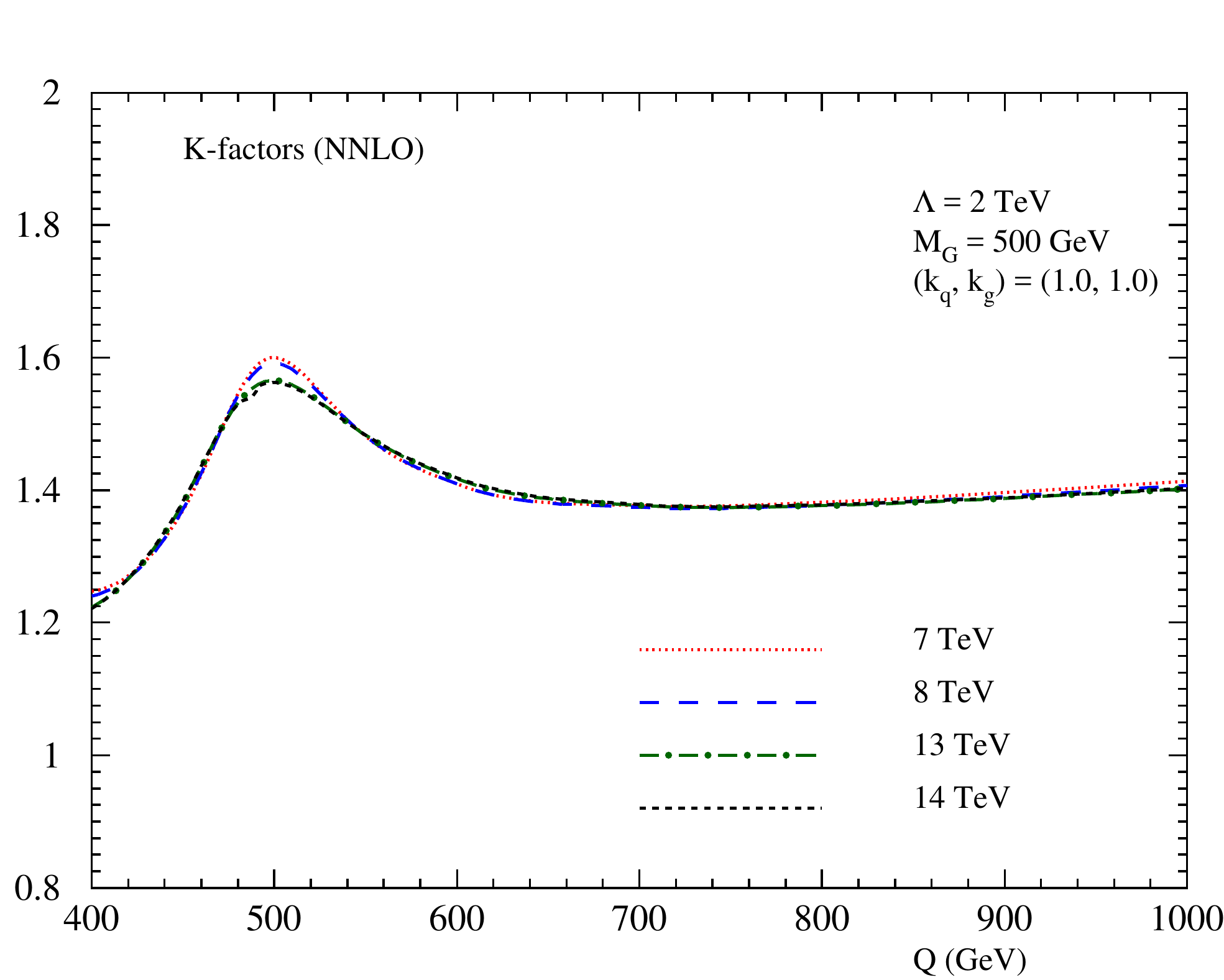,width=7.5cm,height=7.5cm,angle=0}
}
\caption{\sf {Dependence of cross sections on the di-lepton invariant mass
distribution for universal couplings $(1.0,1.0)$.}}
\label{ecm-var1}
\end{figure}

\begin{figure}[htb]
\centerline{
\epsfig{file=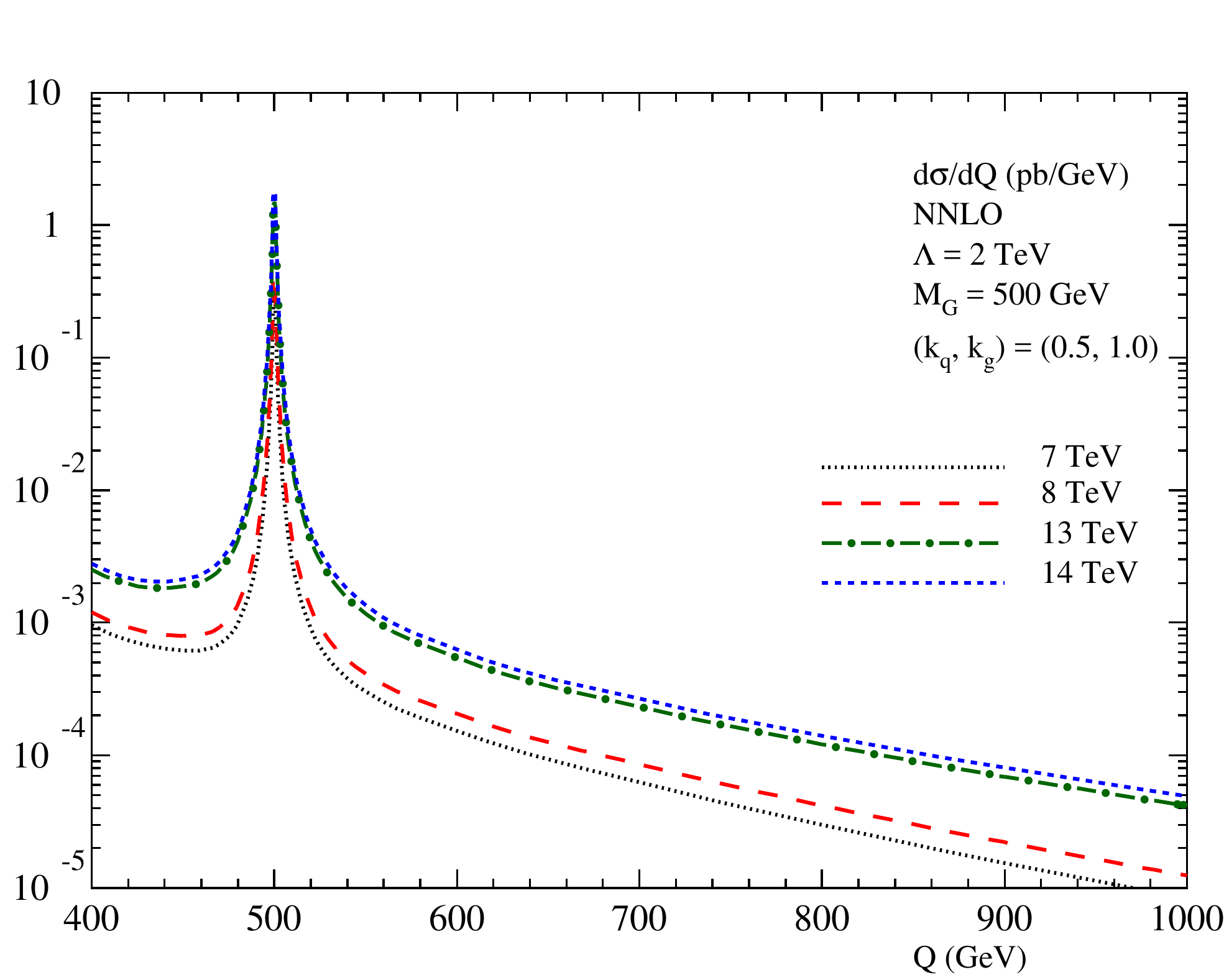,width=7.5cm,height=7.5cm,angle=0}
\epsfig{file=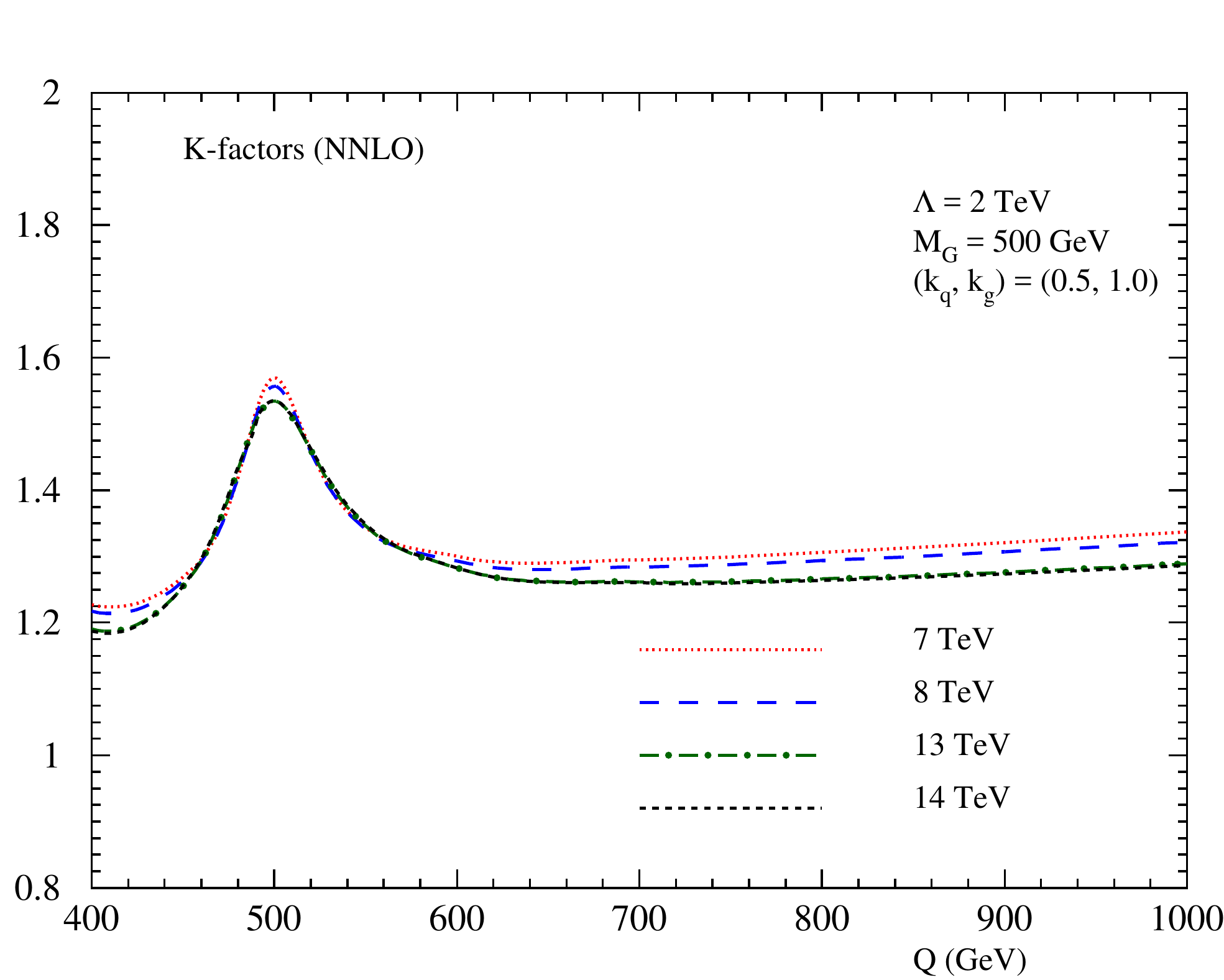,width=7.5cm,height=7.5cm,angle=0}
}
\caption{\sf {Same as fig.\ref{ecm-var1} but for the default choice of
non-universal couplings $(0.5,1.0)$.}}
\label{ecm-var2}
\end{figure}

In what follows, we study the renormalization scale $\mu_R$ and 
the factorization scale $\mu_F$ uncertainties in our predictions.
For this, we define the ratios $R(\mu_R,\mu_F)$ of the invariant mass
distributions computed at arbitrary scale to those computed at the
fixed scale. These are defined as
\begin{eqnarray*}
\text{R}(\mu_R,\mu_F) = \frac{d\sigma(\mu_R, \mu_F)/dQ}{d\sigma(Q_0,Q_0)/dQ}.
\end{eqnarray*}
For a systematic study of these scale uncertainties, we use LO (NLO and NNLO) PDFs
for LO (NLO and NNLO) cross sections respectively.
For convenience, we will study at the resonance region i.e. $Q=M=500$ GeV.
The fixed scale is set equal to $Q_0 = M$. In the left panel of fig.\ref{mu-var},
we present $\text{R}(\mu_R,Q_0)$ by varying $\mu_R$ from $0.1 Q$ to $10 Q$ and keeping
$\mu_F=Q_0$ fixed. At LO, there is no scale $\mu_R$ entering the cross section.
The corresponding scale uncertainties at NLO and NNLO are respectively,
about $19\%$ and $5\%$.

In the right panel of fig.\ref{mu-var}, we present $\text{R}(Q_0,\mu_F)$ by
varying $\mu_F$ from $0.1 Q$ to $10 Q$ and keeping $\mu_R=Q_0$ fixed. For this
range of factorization scale variation, the uncertainties in the distributions
at LO, NLO and NNLO are respectively about $49\%$, $31\%$ and $26\%$.

Finally, we present $\text{R}(\mu,\mu)$ (where $\mu_R=\mu_F =\mu$) in fig.~\ref{murf-var} by
varying $\mu$ from $0.1 Q$ to $10 Q$. The corresponding scale uncertainties
at LO, NLO and NNLO are respectively about $49\%$, $52\%$ and $30\%$.

\begin{figure}[htb]
\centerline{
\epsfig{file=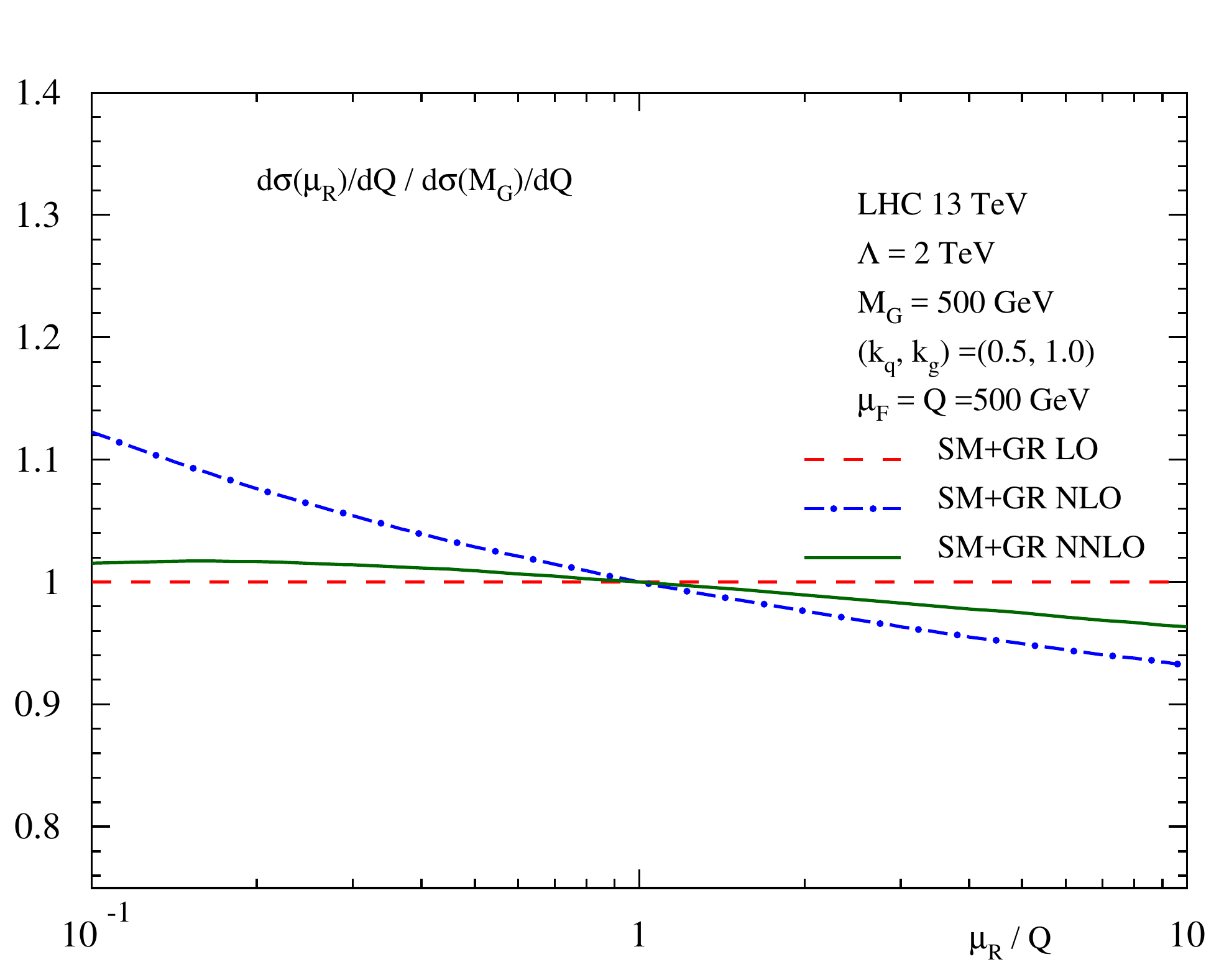,width=7.5cm,height=7.5cm,angle=0}
\epsfig{file=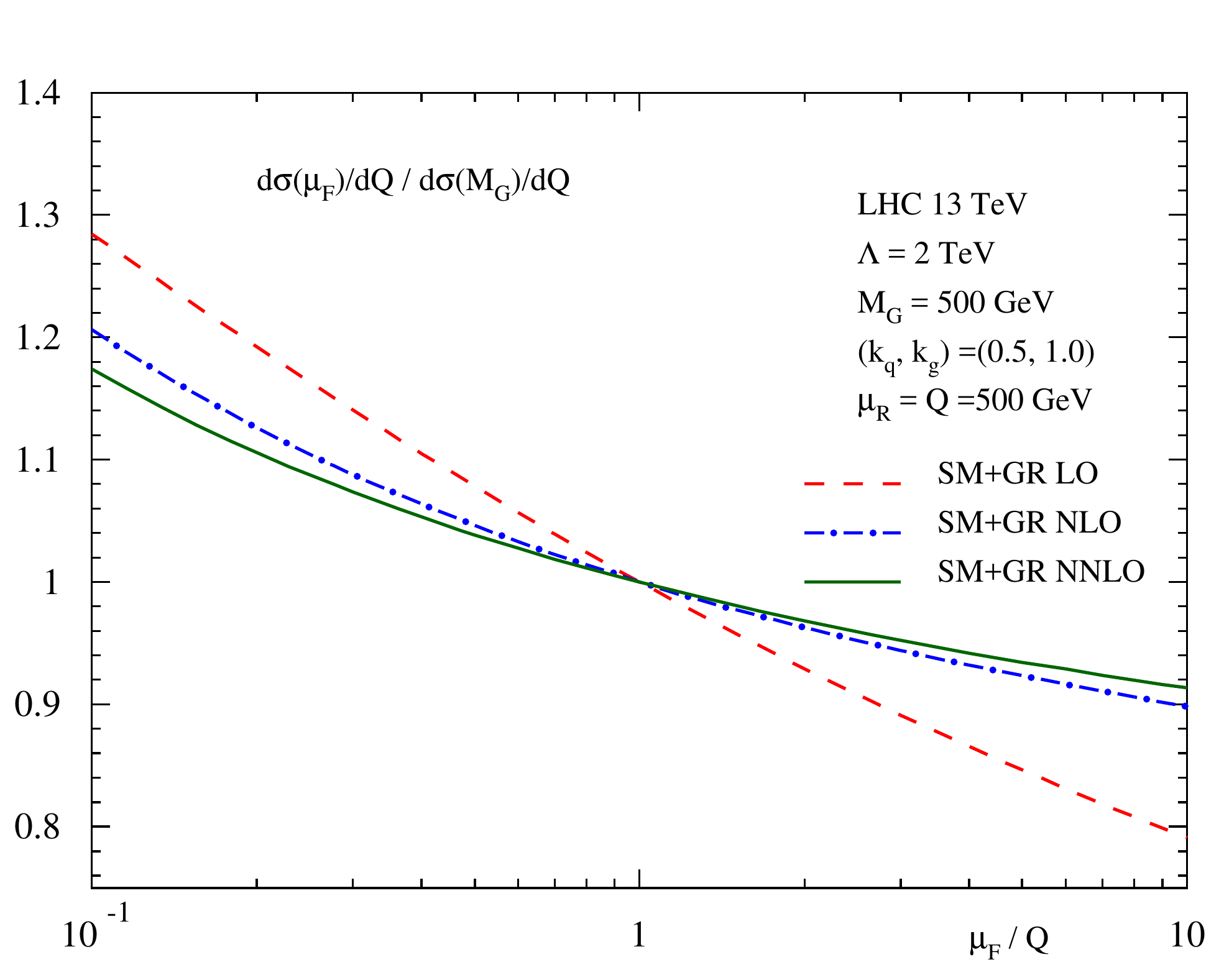,width=7.5cm,height=7.5cm,angle=0}
}
\caption{\sf {Renormalization (left) and factorization (right) scale dependence 
of the di-lepton invariant mass distribution at LO, NLO and NNLO.} }
\label{mu-var}
\end{figure}

\begin{figure}[htb]
\centerline{
\epsfig{file=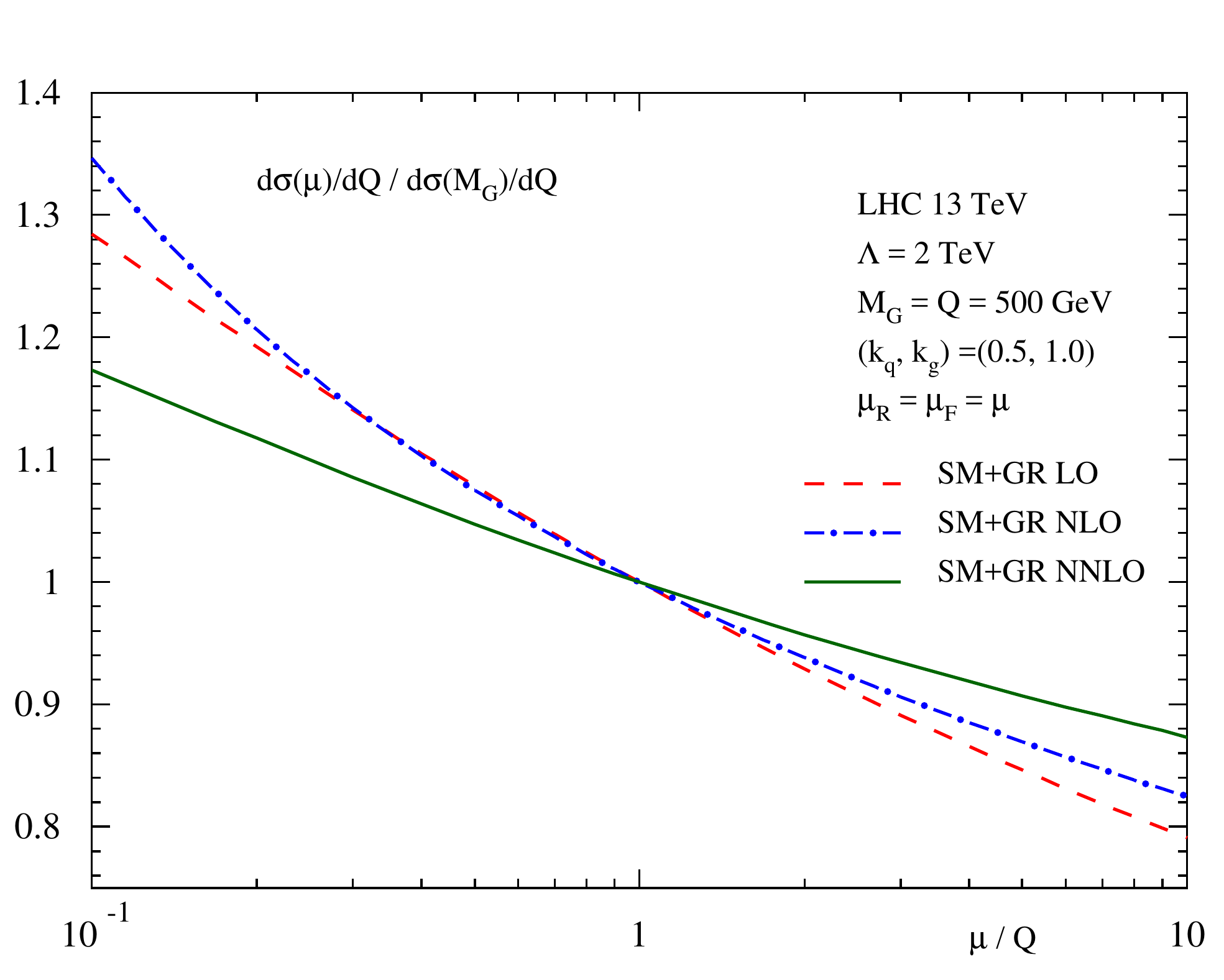,width=7.5cm,height=7.5cm,angle=0}
}
\caption{\sf Same as fig.\ref{mu-var} but with $\mu_R=\mu_F=\mu$.}
\label{murf-var}
\end{figure}

Before we summarize, we also study the uncertainties in our predictions
due to different choice of PDFs used in the calculation. For this analysis,
we make predictions using {\tt MSTW2008, CT10, NNPDF3.0} and {\tt ABM12}
PDFs. The results for the invariant mass distributions for the signal at 
NNLO are presented in the left panel of fig.\ref{pdf-var} and the corresponding 
K-factors are presented in the right panel of fig.\ref{pdf-var}. The K-factors
here are  found to vary from $1.18$ at $Q=400$ GeV to about $1.28$ at $Q=1000$ GeV,
while at the resonance they are about $1.54$.

\begin{figure}[htb]
\centerline{
\epsfig{file=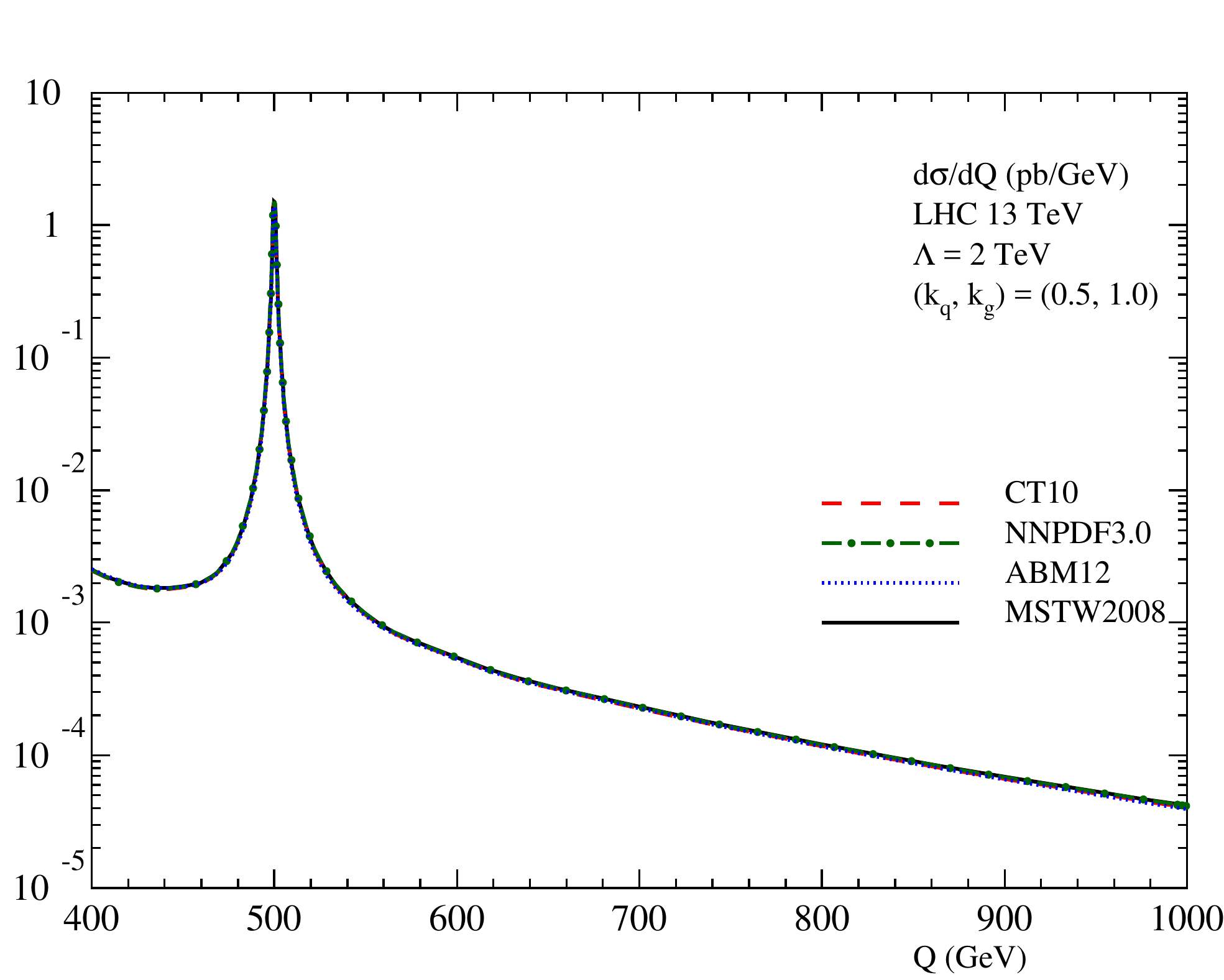,width=7.5cm,height=7.5cm,angle=0}
\epsfig{file=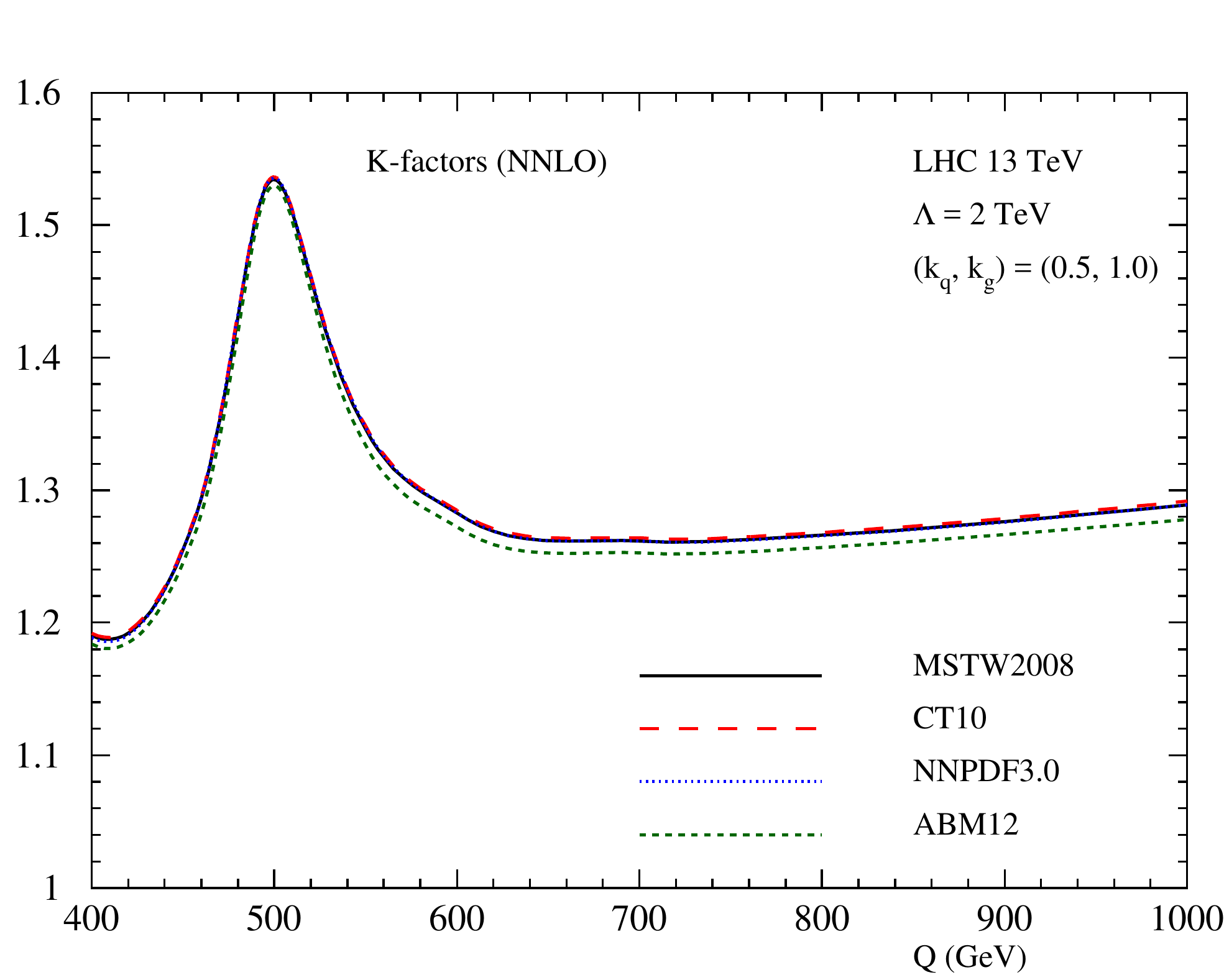,width=7.5cm,height=7.5cm,angle=0}
}
\caption{\sf {Di-lepton invariant mass distributions for different choice
parton distribution functions (PDFs)}. }
\label{pdf-var}
\end{figure}

%\begin{figure}[htb]
%\centerline{
%\epsfig{file=lhc13.c0var.qg.nnlo.cs.pdf,width=7.5cm,height=7.5cm,angle=0}
%}
%\caption{\sf {}}
%\label{cs-qgnnlo}
%\end{figure}
%
\end{section}

\section{Conclusion}
In this article, we have studied for the first time 
the impact of NNLO QCD corrections
to the production of a pair of leptons in the presence of a massive spin-2 
particle at the LHC.
This is done in a minimal scenario where spin-2 particles couple differently to SM
fermions and SM bosons.
This task has been achieved by using the universal IR structure of QCD amplitudes
and the additional UV renormalization that is particularly required for the
case of non-universal couplings, thanks to the recent computations
of the form factors in QCD beyond leading order with non-universal couplings.

Unlike the models with universal couplings, here the 
phenomenology is rich and different. For collider phenomenology at the LHC, we 
present the results for the di-lepton production via spin-2 particle in particular 
for the invariant mass distribution of a pair of leptons for LHC energies. Even at LO, one
can notice that the signal has different cross sections at the resonance region 
in contrast to the gravity mediated models where the signal has the same cross section for
different universal couplings.
At higher orders in QCD, say NLO onwards, the spin-2 exploits its freedom of 
being produced with different coupling strengths even for a given subprocess. This 
particular aspect here makes the QCD radiative corrections crucially dependent
on the choice of the spin-2 coupling strength. Hence the impact of 
QCD corrections here is very much different from those of di-lepton or Higgs production 
in the SM.

We find from our numerical results that the QCD corrections for 
$(k_q,k_g) = (1.0,0.1)$
are dominant over the rest of the choice of couplings, making the K-factors as large
as 2.5 or more. For this choice of couplings, the LO gluon fusion contribution
is very small although gluon fluxes are high for the kinematic region of producing
a 500 GeV particle. But at higher orders where the spin-2 can be emitted off from a 
quark line with large coupling strength, the large quark-gluon fluxes at LHC energies
can potentially enhance the spin-2 production rate, as is evident from the numerical
results. For di-lepton production the `sign' of $qg$ subprocess is usually negative both
in the SM as well as in the models of universal couplings. But here we note that the 
`sign' of $qg$ subprocess contribution changes with the non-universal couplings and 
for the above choice it is positive.

We also gave predictions for different center of mass energies of the incoming protons
at the LHC and found that the K-factors are larger for 7 TeV case. We further quantified
the renormalization and factorization scale uncertainties. For the variation of the scales 
$\mu_R$ and $\mu_F$ between $0.1 Q$ and $10 Q$, the uncertainties are found to get reduced 
from about 50\% at LO to about 30\% at NNLO. For completeness, we also quantified the 
uncertainty in our predictions due to different choice of the PDFs.

These NNLO QCD predictions for the hadroproduction of a massive spin-2 with non-universal
couplings will augment the similar results previously computed at NLO level and compliment
the earlier results for NNLO QCD corrections in models with spin-2/graviton universal couplings.

\section*{Acknowledgement}

We thank G. Das for helping us to validate a part of the numerical results. We also thank T.\ Ahmed and N.\ Rana for useful discussions.
V.\ Ravindran would like to thank M.\ Neubert for the visit at University
of Mainz where the last part of the work was carried out.

\newpage
\appendix
\section{Renormalized form factors}
We present here the results for the renormalised form factors \cite{Ahmed:2016qjf} 
that are used in the present computation. 
In the colour space, the UV renormalised matrix elements of
composite operators ${\cal O}^I,~ I=G,Q$ between a pair of on-shell partonic states $i=q,g$ and
the vacuum state are expanded in powers of coupling constant $a_s$ as
\begin{equation}
 | \mathcal{M}^{I}_i \rangle = \sum_{n=0}^\infty {a}_s^n 
|\mathcal{M}^{I,(n)}_i \rangle
\end{equation}
where $i=q,\overline q, g$.
The on-shell form factor of $\hat {\cal O}^I,~I=G,Q$ are defined by taking the
the overlap of $| \mathcal{M}^{I}_i \rangle$ with its leading order amplitude normalised
with respect to the leading order contribution. We find that there are  
four independent form factors:
\begin{equation}
 {\cal F}^{I,g,(n)} = \frac{\langle {\mathcal M}^{G,(0)}_g | {\mathcal M}^{I,(n)}_g \rangle}{\langle {\mathcal M}^{G,(0)}_g | {\mathcal M}^{G,(0)}_g \rangle} \,,
 \hspace{1cm}
 {\cal F}^{I,q,(n)} = \frac{\langle {\mathcal M}^{Q,(0)}_q | {\mathcal M}^{I,(n)}_q \rangle}{\langle {\mathcal M}^{Q,(0)}_q | {\mathcal M}^{Q,(0)}_q \rangle} \quad \quad \quad I=G,Q\,.
\end{equation}
Note that, the non-diagonal amplitudes \textit{i.e.} $| {\mathcal M}^{Q,(n)}_g \rangle$ and $| {\mathcal M}^{G,(n)}_q \rangle$, start at one-loop level
and hence, the corresponding form factors start at ${\cal O} ({a}_s)$.  The relevant UV renormalised form factors up to two loop level are given below:

\begin{eqnarray}
   \mathcal {F}^{G,g,(1)} &=&
        {1 \over \eps^2} C_A   \Bigg[
          - 8
          \Bigg]
  + {1 \over \eps}  \Bigg[C_A \Bigg(
           \frac{22}{3}
          \Bigg)
	+ n_f   \Bigg(
          - \frac{4}{3}
	  \Bigg)\Bigg]
       + C_A   \Bigg[
          - \frac{203}{18}
          + \zeta_2
          \Bigg]
\nonumber\\&&
       + \eps C_A   \Bigg[
           \frac{2879}{216}
          - \frac{7}{3} \zeta_3
          - \frac{11}{12} \zeta_2
          \Bigg]
       + \eps^2 C_A   \Bigg[
          - \frac{37307}{2592}
          + \frac{77}{36} \zeta_3
          + \frac{203}{144} \zeta_2
          + \frac{47}{80} \zeta_2^2
          \Bigg]\,,
\nonumber\\
   \mathcal {F}^{G,g,(2)} &=&
        {1 \over \eps^4} C_A^2   \Bigg[
           32
          \Bigg]
  + {1 \over \eps^3}    \Bigg[C_A^2\Bigg(
          - \frac{308}{3}
	  \Bigg)
	+ n_f C_A   \Bigg(
           \frac{56}{3}
	  \Bigg)\Bigg]
	+ {1 \over \eps^2}   \Bigg[C_A^2\Bigg(
           \frac{1162}{9}
          - 4 \zeta_2
	  \Bigg)
\nonumber\\&&
	+ n_f C_A   \Bigg(
          - \frac{52}{3}
	  \Bigg)
	+ n_f^2   \Bigg(
           \frac{16}{9}
	  \Bigg)\Bigg]
	+ {1 \over \eps}   \Bigg[C_A^2\Bigg(
          - \frac{4420}{27}
          + \frac{50}{3} \zeta_3
          + 11 \zeta_2
	  \Bigg)
	  \nonumber \\&&
	+ n_f C_A   \Bigg(
           \frac{278}{27}
          - \frac{2}{3} \zeta_2
	  \Bigg)
	+ n_f C_F   \Bigg(
          - 2
	  \Bigg)\Bigg]
       + C_A^2   \Bigg(
           \frac{11854}{81}
          - \frac{253}{9} \zeta_3
          - \frac{94}{9} \zeta_2
\nonumber\\&&
          - \frac{21}{5} \zeta_2^2
	  \Bigg)
       + n_f C_A   \Bigg(
           \frac{1105}{162}
          + 2 \zeta_3
          - \frac{26}{9} \zeta_2
	  \Bigg)
       + n_f C_F   \Bigg(
           \frac{1049}{162}
          - \frac{16}{9} \zeta_2
	  \Bigg)\,,
\nonumber\\
   \mathcal {F}^{G,q,(1)} &=&
        C_F   \Bigg(
           \frac{34}{9}
	  \Bigg)
       + \eps C_F   \Bigg(
          - \frac{79}{27}
          + \frac{2}{3} \zeta_2
	  \Bigg)
       + \eps^2 C_F   \Bigg(
           \frac{401}{162}
          - \frac{14}{9} \zeta_3
          - \frac{17}{36} \zeta_2
	  \Bigg)\,,
\nonumber\\
   \mathcal {F}^{G,q,(2)} &=&
        {1 \over \eps^2} C_F^2   \Bigg(
          - \frac{272}{9}
	  \Bigg)
       + {1 \over \eps} C_F^2   \Bigg(
           \frac{1244}{27}
          - \frac{16}{3} \zeta_2
	  \Bigg)
       + C_F C_A   \Bigg(
           \frac{3913}{81}
          + \frac{28}{9} \zeta_2
	  \Bigg)
\nonumber\\&&
       + C_F^2   \Bigg(
          - \frac{2603}{27}
          + \frac{112}{9} \zeta_3
          + \frac{56}{3} \zeta_2
	  \Bigg)
       + n_f C_F   \Bigg(
          - \frac{1166}{81}
	  \Bigg)\,,
\nonumber\\
   \mathcal {F}^{Q,g,(1)} &=&
        n_f   \Bigg(
           \frac{35}{18}
	  \Bigg)
       + \eps n_f   \Bigg(
          - \frac{497}{216}
          + \frac{1}{6} \zeta_2
	  \Bigg)
       + \eps^2 n_f   \Bigg(
           \frac{6593}{2592}
          - \frac{7}{18} \zeta_3
          - \frac{35}{144} \zeta_2
	  \Bigg)\,,
\nonumber\\
   \mathcal{F}^{Q,g,(2)} &=&
        {1 \over \eps^2} n_f C_A   \Bigg(
          - \frac{140}{9}
	  \Bigg)
       + {1 \over \eps}  \Bigg[n_f C_A\Bigg(
           \frac{98}{3}
          - \frac{4}{3} \zeta_2
	  \Bigg)
       + n_f^2   \Bigg(
          - \frac{70}{27}
	  \Bigg)\Bigg]
\nonumber\\&&
       + n_f C_A   \Bigg(
          - \frac{7625}{162}
          + \frac{100}{9} \zeta_3
          + \frac{53}{9} \zeta_2
	  \Bigg)
       + n_f C_F   \Bigg(
           \frac{299}{81}
          - 8 \zeta_3
          + \frac{16}{9} \zeta_2
	  \Bigg)
\nonumber\\&&
       + n_f^2   \Bigg(
           \frac{497}{162}
          - \frac{2}{9} \zeta_2
	  \Bigg)\,,
\nonumber\\
   \mathcal {F}^{Q,q,(1)} &=&
        {1 \over \eps^2} C_F   \Bigg(
          - 8
	  \Bigg)
       + {1 \over \eps} C_F   \Bigg(
           6
	  \Bigg)
       + C_F   \Bigg(
          - \frac{124}{9}
          + \zeta_2
	  \Bigg)
\nonumber\\&&
       + \eps C_F   \Bigg(
           \frac{403}{27}
          - \frac{7}{3} \zeta_3
          - \frac{17}{12} \zeta_2
	  \Bigg)
       + \eps^2 C_F   \Bigg(
          - \frac{2507}{162}
          + \frac{119}{36} \zeta_3
          + \frac{31}{18} \zeta_2
          + \frac{47}{80} \zeta_2^2
	  \Bigg)\,,
\nonumber\\
   \mathcal {F}^{Q,q,(2)} &=&
        {1 \over \eps^4} C_F^2   \Bigg(
           32
	  \Bigg)
       + {1 \over \eps^3} \Bigg[C_F C_A\Bigg(
          - 44
	  \Bigg)
       + C_F^2   \Bigg(
          - 48
	  \Bigg)
\nonumber\\&&
       +  n_f C_F   \Bigg(
           8
	  \Bigg)\Bigg]
       + {1 \over \eps^2} \Bigg[C_F C_A\Bigg(
           \frac{64}{9}
          + 4 \zeta_2
	  \Bigg)
       + C_F^2   \Bigg(
           \frac{1154}{9}
          - 8 \zeta_2
	  \Bigg)
\nonumber\\&&
       + n_f C_F   \Bigg(
          - \frac{16}{9}
	  \Bigg)\Bigg]
       + {1 \over \eps}  \Bigg[C_F C_A\Bigg(
           \frac{961}{54}
          - 26 \zeta_3
          + 11 \zeta_2
	  \Bigg)
       + C_F^2   \Bigg(
          - \frac{10831}{54}
\nonumber\\&&
          + \frac{128}{3} \zeta_3
          + \frac{16}{3} \zeta_2
	  \Bigg)
       + n_f C_F   \Bigg(
          - \frac{65}{27}
          - 2 \zeta_2
	  \Bigg)\Bigg]
       + C_F C_A   \Bigg(
          - \frac{31495}{216}
          + \frac{601}{9} \zeta_3
\nonumber\\&&
          - \frac{73}{2} \zeta_2
          + \frac{44}{5} \zeta_2^2
	  \Bigg)
       + C_F^2   \Bigg(
           \frac{68677}{216}
          - \frac{922}{9} \zeta_3
          + \frac{11}{6} \zeta_2
          - 13 \zeta_2^2
	  \Bigg)
\nonumber\\&&
       + n_f C_F   \Bigg(
           \frac{9469}{324}
          + \frac{2}{9} \zeta_3
          + \frac{47}{9} \zeta_2
	 \Bigg) 
	  \Bigg]\,.
\end{eqnarray}

\bibliography{main_nonuniv_final} 

\providecommand{\href}[2]{#2}\begingroup\raggedright\begin{thebibliography}{10}

\bibitem{Aad:2012tfa}
{\scshape ATLAS} collaboration, G.~Aad et~al., \emph{{Observation of a new
  particle in the search for the Standard Model Higgs boson with the ATLAS
  detector at the LHC}},
  \href{http://dx.doi.org/10.1016/j.physletb.2012.08.020}{\emph{Phys. Lett.}
  {\bf B716} (2012) 1--29}, [\href{http://arxiv.org/abs/1207.7214}{{\tt
  1207.7214}}].

\bibitem{Chatrchyan:2012xdj}
{\scshape CMS} collaboration, S.~Chatrchyan et~al., \emph{{Observation of a new
  boson at a mass of 125 GeV with the CMS experiment at the LHC}},
  \href{http://dx.doi.org/10.1016/j.physletb.2012.08.021}{\emph{Phys. Lett.}
  {\bf B716} (2012) 30--61}, [\href{http://arxiv.org/abs/1207.7235}{{\tt
  1207.7235}}].

\bibitem{Harlander:2002wh}
R.~V. Harlander and W.~B. Kilgore, \emph{{Next-to-next-to-leading order Higgs
  production at hadron colliders}},
  \href{http://dx.doi.org/10.1103/PhysRevLett.88.201801}{\emph{Phys. Rev.
  Lett.} {\bf 88} (2002) 201801},
  [\href{http://arxiv.org/abs/hep-ph/0201206}{{\tt hep-ph/0201206}}].

\bibitem{Anastasiou:2002yz}
C.~Anastasiou and K.~Melnikov, \emph{{Higgs boson production at hadron
  colliders in NNLO QCD}},
  \href{http://dx.doi.org/10.1016/S0550-3213(02)00837-4}{\emph{Nucl. Phys.}
  {\bf B646} (2002) 220--256}, [\href{http://arxiv.org/abs/hep-ph/0207004}{{\tt
  hep-ph/0207004}}].

\bibitem{Ravindran:2003um}
V.~Ravindran, J.~Smith and W.~L. van Neerven, \emph{{NNLO corrections to the
  total cross-section for Higgs boson production in hadron hadron collisions}},
  {\emph{Nucl. Phys.} {\bf B665} (2003) 325--366}.

\bibitem{Degrassi:2012ry}
G.~Degrassi, S.~Di~Vita, J.~Elias-Miro, J.~R. Espinosa, G.~F. Giudice,
  G.~Isidori et~al., \emph{{Higgs mass and vacuum stability in the Standard
  Model at NNLO}}, \href{http://dx.doi.org/10.1007/JHEP08(2012)098}{\emph{JHEP}
  {\bf 08} (2012) 098}, [\href{http://arxiv.org/abs/1205.6497}{{\tt
  1205.6497}}].

\bibitem{Randall:1999ee}
L.~Randall and R.~Sundrum, \emph{{A Large mass hierarchy from a small extra
  dimension}}, \href{http://dx.doi.org/10.1103/PhysRevLett.83.3370}{\emph{Phys.
  Rev. Lett.} {\bf 83} (1999) 3370--3373},
  [\href{http://arxiv.org/abs/hep-ph/9905221}{{\tt hep-ph/9905221}}].

\bibitem{Khachatryan:2016yec}
{\scshape CMS} collaboration, V.~Khachatryan et~al., \emph{{Search for
  high-mass diphoton resonances in proton–proton collisions at 13 TeV and
  combination with 8 TeV search}},
  \href{http://dx.doi.org/10.1016/j.physletb.2017.01.027}{\emph{Phys. Lett.}
  {\bf B767} (2017) 147--170}, [\href{http://arxiv.org/abs/1609.02507}{{\tt
  1609.02507}}].

\bibitem{Aaboud:2017eta}
{\scshape ATLAS} collaboration, M.~Aaboud et~al., \emph{{Search for diboson
  resonances with boson-tagged jets in $pp$ collisions at $\sqrt{s}=13$ TeV
  with the ATLAS detector}},
  \href{http://dx.doi.org/10.1016/j.physletb.2017.12.011}{\emph{Phys. Lett.}
  {\bf B777} (2018) 91--113}, [\href{http://arxiv.org/abs/1708.04445}{{\tt
  1708.04445}}].

\bibitem{Artoisenet:2013puc}
P.~Artoisenet et~al., \emph{{A framework for Higgs characterisation}},
  \href{http://dx.doi.org/10.1007/JHEP11(2013)043}{\emph{JHEP} {\bf 11} (2013)
  043}, [\href{http://arxiv.org/abs/1306.6464}{{\tt 1306.6464}}].

\bibitem{Ahmed:2016qjf}
T.~Ahmed, P.~Banerjee, P.~K. Dhani, P.~Mathews, N.~Rana and V.~Ravindran,
  \emph{{Three loop form factors of a massive spin-2 particle with nonuniversal
  coupling}}, \href{http://dx.doi.org/10.1103/PhysRevD.95.034035}{\emph{Phys.
  Rev.} {\bf D95} (2017) 034035}, [\href{http://arxiv.org/abs/1612.00024}{{\tt
  1612.00024}}].

\bibitem{Nielsen:1977sy}
N.~K. Nielsen, \emph{{The Energy Momentum Tensor in a Nonabelian Quark Gluon
  Theory}}, \href{http://dx.doi.org/10.1016/0550-3213(77)90040-2}{\emph{Nucl.
  Phys.} {\bf B120} (1977) 212--220}.

\bibitem{Mathews:2004xp}
P.~Mathews, V.~Ravindran, K.~Sridhar and W.~L. van Neerven,
  \emph{{Next-to-leading order QCD corrections to the Drell-Yan cross section
  in models of TeV-scale gravity}},
  \href{http://dx.doi.org/10.1016/j.nuclphysb.2005.01.051}{\emph{Nucl. Phys.}
  {\bf B713} (2005) 333--377}, [\href{http://arxiv.org/abs/hep-ph/0411018}{{\tt
  hep-ph/0411018}}].

\bibitem{Mathews:2005zs}
P.~Mathews and V.~Ravindran, \emph{{Angular distribution of Drell-Yan process
  at hadron colliders to NLO-QCD in models of TeV scale gravity}},
  \href{http://dx.doi.org/10.1016/j.nuclphysb.2006.06.039}{\emph{Nucl. Phys.}
  {\bf B753} (2006) 1--15}, [\href{http://arxiv.org/abs/hep-ph/0507250}{{\tt
  hep-ph/0507250}}].

\bibitem{Kumar:2006id}
M.~C. Kumar, P.~Mathews and V.~Ravindran, \emph{{PDF and scale uncertainties of
  various DY distributions in ADD and RS models at hadron colliders}},
  \href{http://dx.doi.org/10.1140/epjc/s10052-006-0054-0}{\emph{Eur. Phys. J.}
  {\bf C49} (2007) 599--611}, [\href{http://arxiv.org/abs/hep-ph/0604135}{{\tt
  hep-ph/0604135}}].

\bibitem{Kumar:2008pk}
M.~C. Kumar, P.~Mathews, V.~Ravindran and A.~Tripathi, \emph{{Diphoton signals
  in theories with large extra dimensions to NLO QCD at hadron colliders}},
  \href{http://dx.doi.org/10.1016/j.physletb.2009.01.002}{\emph{Phys. Lett.}
  {\bf B672} (2009) 45--50}, [\href{http://arxiv.org/abs/0811.1670}{{\tt
  0811.1670}}].

\bibitem{Kumar:2009nn}
M.~C. Kumar, P.~Mathews, V.~Ravindran and A.~Tripathi, \emph{{Direct photon
  pair production at the LHC to order $\alpha_s$ in TeV scale gravity models}},
  \href{http://dx.doi.org/10.1016/j.nuclphysb.2009.03.022}{\emph{Nucl. Phys.}
  {\bf B818} (2009) 28--51}, [\href{http://arxiv.org/abs/0902.4894}{{\tt
  0902.4894}}].

\bibitem{Agarwal:2009xr}
N.~Agarwal, V.~Ravindran, V.~K. Tiwari and A.~Tripathi, \emph{{Z boson pair
  production at the LHC to O(alpha(s)) in TeV scale gravity models}},
  \href{http://dx.doi.org/10.1016/j.nuclphysb.2009.12.032}{\emph{Nucl. Phys.}
  {\bf B830} (2010) 248--270}, [\href{http://arxiv.org/abs/0909.2651}{{\tt
  0909.2651}}].

\bibitem{Agarwal:2009zg}
N.~Agarwal, V.~Ravindran, V.~K. Tiwari and A.~Tripathi, \emph{{Next-to-leading
  order QCD corrections to the $Z$ boson pair production at the LHC in Randall
  Sundrum model}},
  \href{http://dx.doi.org/10.1016/j.physletb.2010.02.060}{\emph{Phys. Lett.}
  {\bf B686} (2010) 244--248}, [\href{http://arxiv.org/abs/0910.1551}{{\tt
  0910.1551}}].

\bibitem{Agarwal:2010sp}
N.~Agarwal, V.~Ravindran, V.~K. Tiwari and A.~Tripathi, \emph{{$W^+W^-$
  production in Large extra dimension model at next-to-leading order in QCD at
  the LHC}}, \href{http://dx.doi.org/10.1103/PhysRevD.82.036001}{\emph{Phys.
  Rev.} {\bf D82} (2010) 036001}, [\href{http://arxiv.org/abs/1003.5450}{{\tt
  1003.5450}}].

\bibitem{Agarwal:2010sn}
N.~Agarwal, V.~Ravindran, V.~K. Tiwari and A.~Tripathi, \emph{{Next-to-leading
  order QCD corrections to $W^+W^-$ production at the LHC in Randall Sundrum
  model}}, \href{http://dx.doi.org/10.1016/j.physletb.2010.05.063}{\emph{Phys.
  Lett.} {\bf B690} (2010) 390--395},
  [\href{http://arxiv.org/abs/1003.5445}{{\tt 1003.5445}}].

\bibitem{Frederix:2012dp}
R.~Frederix, M.~K. Mandal, P.~Mathews, V.~Ravindran, S.~Seth, P.~Torrielli
  et~al., \emph{{Diphoton production in the ADD model to NLO+parton shower
  accuracy at the LHC}},
  \href{http://dx.doi.org/10.1007/JHEP12(2012)102}{\emph{JHEP} {\bf 12} (2012)
  102}, [\href{http://arxiv.org/abs/1209.6527}{{\tt 1209.6527}}].

\bibitem{Frederix:2013lga}
R.~Frederix, M.~K. Mandal, P.~Mathews, V.~Ravindran and S.~Seth,
  \emph{{Drell-Yan, $ZZ, W^+W^-$ production in SM \& ADD model to NLO+PS
  accuracy at the LHC}},
  \href{http://dx.doi.org/10.1140/epjc/s10052-014-2745-2}{\emph{Eur. Phys. J.}
  {\bf C74} (2014) 2745}, [\href{http://arxiv.org/abs/1307.7013}{{\tt
  1307.7013}}].

\bibitem{Das:2014tva}
G.~Das, P.~Mathews, V.~Ravindran and S.~Seth, \emph{{RS resonance in di-final
  state production at the LHC to NLO+PS accuracy}},
  \href{http://dx.doi.org/10.1007/JHEP10(2014)188}{\emph{JHEP} {\bf 10} (2014)
  188}, [\href{http://arxiv.org/abs/1408.3970}{{\tt 1408.3970}}].

\bibitem{Das:2016pbk}
G.~Das, C.~Degrande, V.~Hirschi, F.~Maltoni and H.-S. Shao, \emph{{NLO
  predictions for the production of a (750 GeV) spin-two particle at the LHC}},
   \href{http://arxiv.org/abs/1605.09359}{{\tt 1605.09359}}.

\bibitem{deFlorian:2013sza}
D.~de~Florian, M.~Mahakhud, P.~Mathews, J.~Mazzitelli and V.~Ravindran,
  \emph{{Quark and gluon spin-2 form factors to two-loops in QCD}},
  \href{http://dx.doi.org/10.1007/JHEP02(2014)035}{\emph{JHEP} {\bf 02} (2014)
  035}, [\href{http://arxiv.org/abs/1312.6528}{{\tt 1312.6528}}].

\bibitem{deFlorian:2013wpa}
D.~de~Florian, M.~Mahakhud, P.~Mathews, J.~Mazzitelli and V.~Ravindran,
  \emph{{Next-to-Next-to-Leading Order QCD Corrections in Models of TeV-Scale
  Gravity}}, \href{http://dx.doi.org/10.1007/JHEP04(2014)028}{\emph{JHEP} {\bf
  04} (2014) 028}, [\href{http://arxiv.org/abs/1312.7173}{{\tt 1312.7173}}].

\bibitem{Ahmed:2016qhu}
T.~Ahmed, P.~Banerjee, P.~K. Dhani, M.~C. Kumar, P.~Mathews, N.~Rana et~al.,
  \emph{{NNLO QCD Corrections to the Drell-Yan Cross Section in Models of
  TeV-Scale Gravity}},  \href{http://arxiv.org/abs/1606.08454}{{\tt
  1606.08454}}.

\bibitem{Ahmed:2014gla}
T.~Ahmed, M.~Mahakhud, P.~Mathews, N.~Rana and V.~Ravindran, \emph{{Two-Loop
  QCD Correction to massive spin-2 resonance $\rightarrow$ 3 gluons}},
  \href{http://dx.doi.org/10.1007/JHEP05(2014)107}{\emph{JHEP} {\bf 05} (2014)
  107}, [\href{http://arxiv.org/abs/1404.0028}{{\tt 1404.0028}}].

\bibitem{Ahmed:2016yox}
T.~Ahmed, G.~Das, P.~Mathews, N.~Rana and V.~Ravindran, \emph{{The two-loop QCD
  correction to massive spin-2 resonance $ \rightarrow q \bar{q} g $}},
  \href{http://dx.doi.org/10.1140/epjc/s10052-016-4478-x}{\emph{Eur. Phys. J.}
  {\bf C76} (2016) 667}, [\href{http://arxiv.org/abs/1608.05906}{{\tt
  1608.05906}}].

\bibitem{Aad:2015mxa}
{\scshape ATLAS} collaboration, G.~Aad et~al., \emph{{Study of the spin and
  parity of the Higgs boson in diboson decays with the ATLAS detector}},
  \href{http://dx.doi.org/10.1140/epjc/s10052-015-3685-1,
  10.1140/epjc/s10052-016-3934-y}{\emph{Eur. Phys. J.} {\bf C75} (2015) 476},
  [\href{http://arxiv.org/abs/1506.05669}{{\tt 1506.05669}}].

\bibitem{Pedersen:2015jdh}
L.~E. Pedersen, \emph{{Probing the nature of the Higgs Boson: A study of the
  Higgs spin and parity through the $ZZ^* \to 4l$ final state at the ATLAS
  Experiment}}.
\newblock PhD thesis, Bohr Inst., 2015.

\bibitem{Han:1998sg}
T.~Han, J.~D. Lykken and R.-J. Zhang, \emph{{On Kaluza-Klein states from large
  extra dimensions}},
  \href{http://dx.doi.org/10.1103/PhysRevD.59.105006}{\emph{Phys. Rev.} {\bf
  D59} (1999) 105006}, [\href{http://arxiv.org/abs/hep-ph/9811350}{{\tt
  hep-ph/9811350}}].

\bibitem{Mathews:2004pi}
P.~Mathews, V.~Ravindran and K.~Sridhar, \emph{{NLO - QCD corrections to e+ e-
  ---> hadrons in models of TeV-scale gravity}},
  \href{http://dx.doi.org/10.1088/1126-6708/2004/08/048}{\emph{JHEP} {\bf 08}
  (2004) 048}, [\href{http://arxiv.org/abs/hep-ph/0405292}{{\tt
  hep-ph/0405292}}].

\bibitem{Altarelli:1978id}
G.~Altarelli, R.~K. Ellis and G.~Martinelli, \emph{{Leptoproduction and
  Drell-Yan Processes Beyond the Leading Approximation in Chromodynamics}},
  \href{http://dx.doi.org/10.1016/0550-3213(78)90085-8,
  10.1016/0550-3213(78)90067-6}{\emph{Nucl. Phys.} {\bf B143} (1978) 521}.

\bibitem{Matsuura:1987wt}
T.~Matsuura and W.~L. van Neerven, \emph{{Second Order Logarithmic Corrections
  to the {Drell-Yan} Cross-section}},
  \href{http://dx.doi.org/10.1007/BF01624369}{\emph{Z. Phys.} {\bf C38} (1988)
  623}.

\bibitem{Matsuura:1988sm}
T.~Matsuura, S.~C. van~der Marck and W.~L. van Neerven, \emph{{The Calculation
  of the Second Order Soft and Virtual Contributions to the Drell-Yan
  Cross-Section}},
  \href{http://dx.doi.org/10.1016/0550-3213(89)90620-2}{\emph{Nucl. Phys.} {\bf
  B319} (1989) 570--622}.

\bibitem{Hamberg:1990np}
R.~Hamberg, W.~L. van Neerven and T.~Matsuura, \emph{{A complete calculation of
  the order $\alpha-s^{2}$ correction to the Drell-Yan $K$ factor}},
  \href{http://dx.doi.org/10.1016/S0550-3213(02)00814-3,
  10.1016/0550-3213(91)90064-5}{\emph{Nucl. Phys.} {\bf B359} (1991) 343--405}.

\bibitem{Ahmed:2014cla}
T.~Ahmed, M.~Mahakhud, N.~Rana and V.~Ravindran, \emph{{Drell-Yan Production at
  Threshold to Third Order in QCD}},
  \href{http://dx.doi.org/10.1103/PhysRevLett.113.112002}{\emph{Phys. Rev.
  Lett.} {\bf 113} (2014) 112002}, [\href{http://arxiv.org/abs/1404.0366}{{\tt
  1404.0366}}].

\bibitem{Tkachov:1981wb}
F.~V. Tkachov, \emph{{A Theorem on Analytical Calculability of Four Loop
  Renormalization Group Functions}},
  \href{http://dx.doi.org/10.1016/0370-2693(81)90288-4}{\emph{Phys. Lett.} {\bf
  B100} (1981) 65--68}.

\bibitem{Chetyrkin:1981qh}
K.~Chetyrkin and F.~Tkachov, \emph{{Integration by Parts: The Algorithm to
  Calculate beta Functions in 4 Loops}},
  \href{http://dx.doi.org/10.1016/0550-3213(81)90199-1}{\emph{Nucl.Phys.} {\bf
  B192} (1981) 159--204}.

\bibitem{Gehrmann:1999as}
T.~Gehrmann and E.~Remiddi, \emph{{Differential equations for two loop four
  point functions}},
  \href{http://dx.doi.org/10.1016/S0550-3213(00)00223-6}{\emph{Nucl.Phys.} {\bf
  B580} (2000) 485--518}.

\bibitem{Anastasiou:2014vaa}
C.~Anastasiou, C.~Duhr, F.~Dulat, E.~Furlan, T.~Gehrmann, F.~Herzog et~al.,
  \emph{{Higgs boson gluon–fusion production at threshold in N$^3$LO QCD}},
  \href{http://dx.doi.org/10.1016/j.physletb.2014.08.067}{\emph{Phys. Lett.}
  {\bf B737} (2014) 325--328}, [\href{http://arxiv.org/abs/1403.4616}{{\tt
  1403.4616}}].

\bibitem{Anastasiou:2015ema}
C.~Anastasiou, C.~Duhr, F.~Dulat, F.~Herzog and B.~Mistlberger, \emph{{Higgs
  Boson Gluon-Fusion Production in QCD at Three Loops}},
  \href{http://dx.doi.org/10.1103/PhysRevLett.114.212001}{\emph{Phys. Rev.
  Lett.} {\bf 114} (2015) 212001}, [\href{http://arxiv.org/abs/1503.06056}{{\tt
  1503.06056}}].

\bibitem{Anastasiou:2016cez}
C.~Anastasiou, C.~Duhr, F.~Dulat, E.~Furlan, T.~Gehrmann, F.~Herzog et~al.,
  \emph{{High precision determination of the gluon fusion Higgs boson
  cross-section at the LHC}},
  \href{http://dx.doi.org/10.1007/JHEP05(2016)058}{\emph{JHEP} {\bf 05} (2016)
  058}, [\href{http://arxiv.org/abs/1602.00695}{{\tt 1602.00695}}].

\bibitem{Kinoshita:1962ur}
T.~Kinoshita, \emph{{Mass singularities of Feynman amplitudes}},
  \href{http://dx.doi.org/10.1063/1.1724268}{\emph{J. Math. Phys.} {\bf 3}
  (1962) 650--677}.

\bibitem{Lee:1964is}
T.~D. Lee and M.~Nauenberg, \emph{{Degenerate Systems and Mass Singularities}},
  \href{http://dx.doi.org/10.1103/PhysRev.133.B1549}{\emph{Phys. Rev.} {\bf
  133} (1964) B1549--B1562}.

\bibitem{Gross:1973id}
D.~J. Gross and F.~Wilczek, \emph{{Ultraviolet Behavior of Nonabelian Gauge
  Theories}}, \href{http://dx.doi.org/10.1103/PhysRevLett.30.1343}{\emph{Phys.
  Rev. Lett.} {\bf 30} (1973) 1343--1346}.

\bibitem{Politzer:1973fx}
H.~D. Politzer, \emph{{Reliable Perturbative Results for Strong
  Interactions?}},
  \href{http://dx.doi.org/10.1103/PhysRevLett.30.1346}{\emph{Phys. Rev. Lett.}
  {\bf 30} (1973) 1346--1349}.

\bibitem{Caswell:1974gg}
W.~E. Caswell, \emph{{Asymptotic Behavior of Nonabelian Gauge Theories to Two
  Loop Order}}, \href{http://dx.doi.org/10.1103/PhysRevLett.33.244}{\emph{Phys.
  Rev. Lett.} {\bf 33} (1974) 244}.

\bibitem{Tarasov:2013zv}
O.~V. Tarasov and A.~A. Vladimirov, \emph{{Three Loop Calculations in
  Non-Abelian Gauge Theories}},
  \href{http://dx.doi.org/10.1134/S1063779613050043}{\emph{Phys. Part. Nucl.}
  {\bf 44} (2013) 791--802}, [\href{http://arxiv.org/abs/1301.5645}{{\tt
  1301.5645}}].

\bibitem{Larin:1993tp}
S.~A. Larin and J.~A.~M. Vermaseren, \emph{{The Three loop QCD Beta function
  and anomalous dimensions}},
  \href{http://dx.doi.org/10.1016/0370-2693(93)91441-O}{\emph{Phys. Lett.} {\bf
  B303} (1993) 334--336}, [\href{http://arxiv.org/abs/hep-ph/9302208}{{\tt
  hep-ph/9302208}}].

\bibitem{Altarelli:1977zs}
G.~Altarelli and G.~Parisi, \emph{{Asymptotic Freedom in Parton Language}},
  \href{http://dx.doi.org/10.1016/0550-3213(77)90384-4}{\emph{Nucl. Phys.} {\bf
  B126} (1977) 298--318}.

\bibitem{Floratos:1980hm}
E.~G. Floratos, R.~Lacaze and C.~Kounnas, \emph{{Space and Timelike Cut
  Vertices in {QCD} Beyond the Leading Order. 2. The Singlet Sector}},
  \href{http://dx.doi.org/10.1016/0370-2693(81)90016-2}{\emph{Phys. Lett.} {\bf
  98B} (1981) 285--290}.

\bibitem{Floratos:1980hk}
E.~G. Floratos, R.~Lacaze and C.~Kounnas, \emph{{Space and Timelike Cut
  Vertices in QCD Beyond the Leading Order. 1. Nonsinglet Sector}},
  \href{http://dx.doi.org/10.1016/0370-2693(81)90374-9}{\emph{Phys. Lett.} {\bf
  98B} (1981) 89--95}.

\bibitem{Curci:1980uw}
G.~Curci, W.~Furmanski and R.~Petronzio, \emph{{Evolution of Parton Densities
  Beyond Leading Order: The Nonsinglet Case}},
  \href{http://dx.doi.org/10.1016/0550-3213(80)90003-6}{\emph{Nucl. Phys.} {\bf
  B175} (1980) 27--92}.

\bibitem{Moch:2004pa}
S.~Moch, J.~A.~M. Vermaseren and A.~Vogt, \emph{{The Three loop splitting
  functions in QCD: The Nonsinglet case}},
  \href{http://dx.doi.org/10.1016/j.nuclphysb.2004.03.030}{\emph{Nucl. Phys.}
  {\bf B688} (2004) 101--134}, [\href{http://arxiv.org/abs/hep-ph/0403192}{{\tt
  hep-ph/0403192}}].

\bibitem{Vogt:2004mw}
A.~Vogt, S.~Moch and J.~A.~M. Vermaseren, \emph{{The Three-loop splitting
  functions in QCD: The Singlet case}},
  \href{http://dx.doi.org/10.1016/j.nuclphysb.2004.04.024}{\emph{Nucl. Phys.}
  {\bf B691} (2004) 129--181}, [\href{http://arxiv.org/abs/hep-ph/0404111}{{\tt
  hep-ph/0404111}}].

\bibitem{Falkowski:2016glr}
A.~Falkowski and J.~F. Kamenik, \emph{{Diphoton portal to warped gravity}},
  \href{http://dx.doi.org/10.1103/PhysRevD.94.015008}{\emph{Phys. Rev.} {\bf
  D94} (2016) 015008}, [\href{http://arxiv.org/abs/1603.06980}{{\tt
  1603.06980}}].

\end{thebibliography}\endgroup
\bibliographystyle{JHEP}
\end{document}